\documentclass{article}
\pdfoutput=1
\usepackage{spconf,amsmath,graphicx}
\usepackage{amsmath,graphicx}
\usepackage{amsfonts, amssymb, amsthm}
\usepackage[T1]{fontenc}    
\usepackage{url}            
\usepackage{booktabs}       
\usepackage{nicefrac}       
\usepackage{microtype}      
\usepackage{bbm}
\usepackage{cite}

\usepackage{multirow}
\usepackage{epsfig}

\usepackage{color}
\usepackage[font=small,labelfont=bf]{caption}
\theoremstyle{definition}

\newtheorem*{remark*}{Remark}

\usepackage{textcase}

\usepackage[colorlinks,
linkcolor=blue,
anchorcolor=blue,
citecolor=blue
]{hyperref}
\title{T\textsc{\MakeTextLowercase{ensor}}M\textsc{\MakeTextLowercase{ap}}: L\textsc{\MakeTextLowercase{idar-based}} T\textsc{\MakeTextLowercase{opological}} M\textsc{\MakeTextLowercase{apping}} \textsc{\MakeTextLowercase{and}} L\textsc{\MakeTextLowercase{ocalization}}\\ \textsc{\MakeTextLowercase{via}} T\textsc{\MakeTextLowercase{ensor}} D\textsc{\MakeTextLowercase{ecompositions}}\vspace{-10pt}}
\name{\fontsize{10pt}{11.5pt}Sirisha Rambhatla$^\dagger$, Nikos D. Sidiropoulos $^\ddagger$ and Jarvis Haupt $^\dagger$\vspace{-10pt}\thanks{The authors graciously acknowledge support from the DARPA Young Faculty Award, Grant N66001-14-1-4047.}}
\address{\fontsize{10pt}{11.5pt}\selectfont $^\dagger$Dept. of Electrical \& Computer Engineering, University of Minnesota--Twin Cities, Minneapolis, MN 55455\\
\fontsize{10pt}{11.5pt}\selectfont $^\ddagger$Dept. of Electrical \& Computer Engineering, University of Virginia, Charlottesville, Virginia 22904\\
\fontsize{10pt}{11.5pt}\selectfont\tt{rambh002@umn.edu, nikos@virginia.edu, jdhaupt@umn.edu }\vspace{0pt}}

\newcommand{\sr}[1]{{\color{black}{#1}}} 
\begin{document} 
\ninept
\maketitle
\vspace{-20pt}
\begin{abstract}
We propose a technique to develop (and localize in) topological maps from light detection and ranging (Lidar) data. Localizing an autonomous vehicle with respect to a reference map in real-time is crucial for its safe operation. Owing to the rich information provided by Lidar sensors, these are emerging as a promising choice for this task. However, since a Lidar outputs a large amount of data every fraction of a second, it is progressively harder to process the information in real-time. Consequently, current systems have migrated towards faster alternatives at the expense of accuracy. 
To overcome this inherent trade-off between latency and accuracy, we propose a technique to develop topological maps from Lidar data using the orthogonal Tucker3 tensor decomposition.  Our experimental evaluations demonstrate that in addition to achieving a high compression ratio as compared to full data, the proposed technique, \emph{TensorMap}, also accurately detects the position of the vehicle \sr{in a graph-based representation of a map}. We also analyze the robustness of the proposed technique to Gaussian and translational noise, thus initiating explorations into potential applications of tensor decompositions in Lidar data analysis.

\end{abstract}

  \vspace{-14pt}
\section{Introduction}\label{sec:intro}
  \vspace{-6pt}
Autonomous vehicles are gaining significant traction due to the advent of smaller footprint, yet fast processors. One of the major steps in autonomous vehicle navigation is to keep track of the state of the vehicle which, among other things, includes the position of the vehicle with respect to the global frame of reference.  For this, vehicles often employ a wide range of sensors like GPS, cameras and inertial measurement units (IMU). 
However, these sensors usually do not provide the accuracies required to establish safe (and stable) operation. 

The advances in Lidar technology coupled with its increasing affordability have made it the most popular sensor for tracking position with millimeter accuracies. However, the Lidar technology comes with its own set of drawbacks. Each \textit{scan} (the range measurements received by the sensors at different angles of azimuth and elevation) obtained by the Lidar sensor is a point cloud containing millions of data points. Although this data provides very accurate details about the operating environment, the sheer volume of the data  thrown at the processor every fraction of a second, often forces us to choose between speed of operation (latency) and accuracy.  

\begin{figure}[t]
\vspace{-5pt}
  \centering\setlength{\belowcaptionskip}{0pt}
    \begin{tabular}{cc}
         \includegraphics[width=0.18\textwidth]{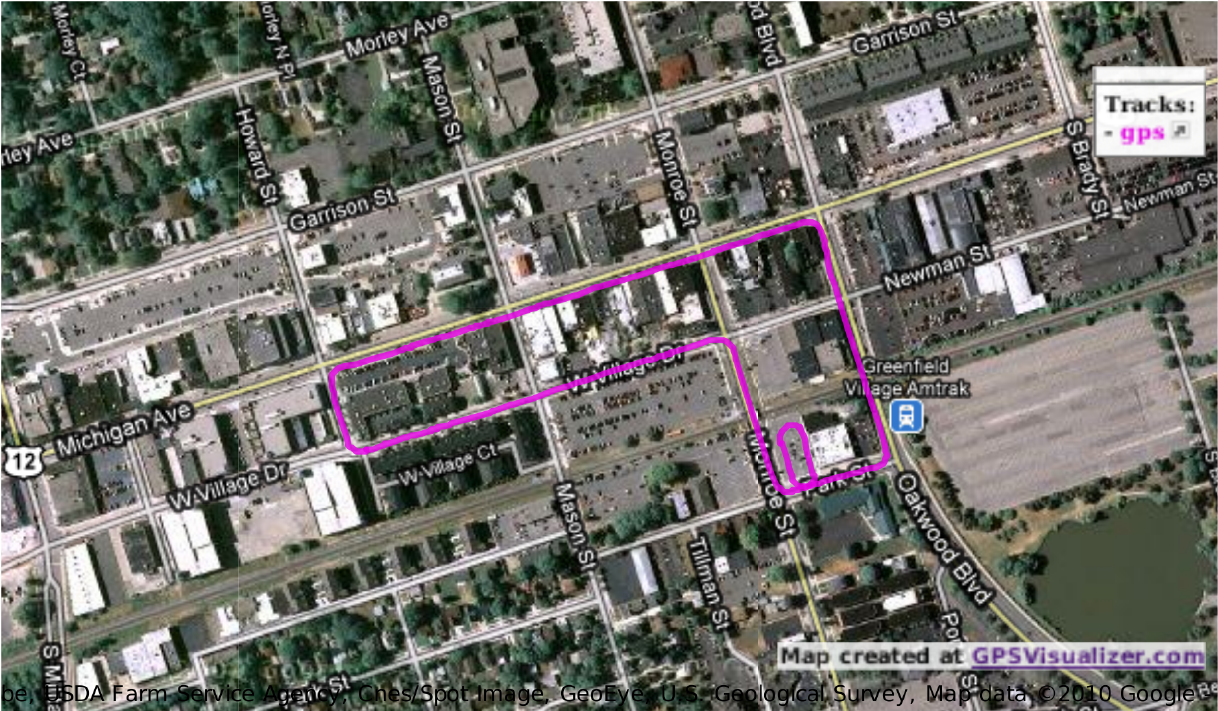} &  \includegraphics[width=0.18\textwidth]{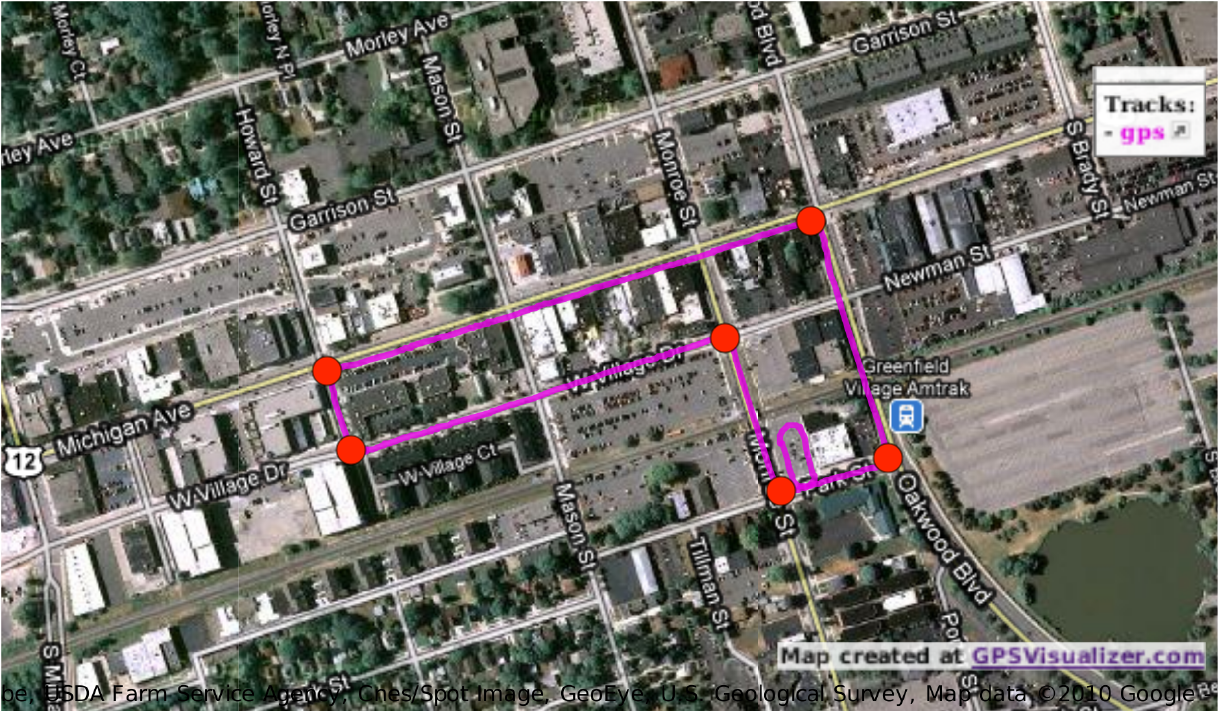}\\
        (a) & (b) 
        \vspace{-8pt}
    \end{tabular}
  \caption{The Ford Dataset \cite{Pandey11}. Panels (a) and (b) show the trajectory traced by the vehicle, and nodes of a representative topological map (in red), respectively. }          \vspace{-20pt}
  \label{fig:map} 
\end{figure}

One way of addressing this issue is to develop efficient representations of the map. To develop these representations, often a map as the one shown in Fig.~\ref{fig:map} (a), can be viewed as a graph with nodes as turns/landmarks, with roads as the edges or segments of the graph. Such a map is known as a \textit{topological map}; Fig.~\ref{fig:map} (b) shows an example of the nodes in such a map. The problem of localization then becomes a problem of identifying which segment the vehicle is on, and how far along in the segment it is positioned. 
\vspace{-12pt}
\subsection{Prior-Art}
\vspace{-6pt}
Building topological maps for localization using imaging-based techniques has gained traction in recent times since these are inexpensive to implement and faster to process \cite{Siagian2009, Wang2006, Fraundorfer2007, Booij2007, Chang2010, Milford2012, Schindler2007, Angeli2009}, as compared to Lidar sensors. \sr{However, these vision-based techniques are sensitive to changing weather and illumination (day and night).} 

\sr{The process of identifying the rigid body transformation that aligns a scan with a map is known as \textit{scan matching}, and is a very effective choice for localization.}
Significant advances have been made in the area of developing better and accurate representations for scan-matching using Lidar data \cite{Besl92, Biber03,Morris2005,Myronenko10, Mueller2011}, but the time, and computational overhead, associated with it are still prohibitive. \sr{The state-of-the-art techniques deal with the computational overhead by acquiring Lidar data at lower rate in order to operate in real-time \cite{zhang14, zhang15}}. 

On the other hand, low rank tensor models, specifically Tucker3 \cite{Tucker66} decomposition, popularized by the higher-order singular value decomposition (HO-SVD) technique\cite{De2000}, have gained success in a wide variety of applications; see \cite{Kolda09, Sidiropoulos2017} and the references therein for details. Viewed as a generalization of SVD, here the tensor is factorized as \textit{core} tensor multiplied by factor matrices in each dimension (mode); the size of the matrices controlling the respective mode ranks (collectively, the so-called multi-linear rank of the tensor). In addition to compressing approximately low multi-linear rank tensors, this decomposition exhibits an interesting property -- the core tensor is \textit{all orthogonal}, i.e., each slice of this tensor is orthogonal to all the other slices; see \cite{De2000} for details. 

It is worth noting that recently, \cite{Li2017} employed tensor models to classify objects in a Lidar scan based on dictionary learning. As opposed to this work, our aim here is to localize a vehicle on a map using the Lidar scans. 

\begin{figure*}[t]
\centering
\includegraphics[width=1\textwidth]{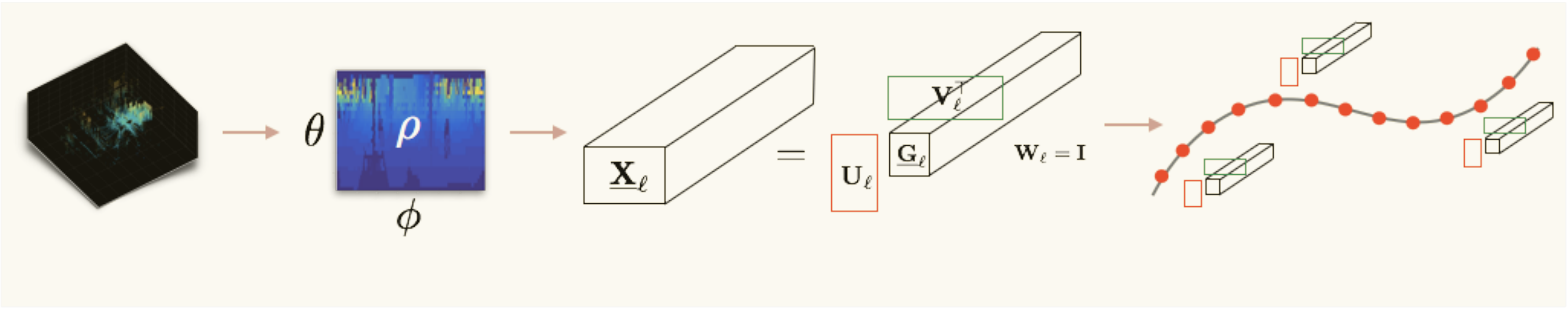} \\\vspace{-25pt}
\hspace{-40pt} (a) $3$-D Point Cloud \hspace{12pt} (b) Matricized scan,   \hspace{15pt} (c) Learn Tucker$3$ models for each length-$k$ \hspace{50pt}(d) TensorMap.\\
 \hspace{-220pt} (a Lidar scan), \hspace{20pt} \vspace{-2pt} \hspace{100pt}segment tensor,\vspace{-3pt}
\caption{Learning the topological map. We represent each $3$-D point cloud corresponding to each Lidar scan (a), as a matrix (b) after conversion to polar coordinates. We aggregate the matricized scans to form length-$k$ segment tensors $\underline{\mathbf{X}}_\ell$, and learn the orthogonal Tucker3 models on each of these (shown in panels (c) and (d)). }
\label{figure:Prep}
  \vspace{-18pt}
\end{figure*}
\begin{figure}
  \centering\setlength{\belowcaptionskip}{0pt}
    \includegraphics[width=0.48\textwidth]{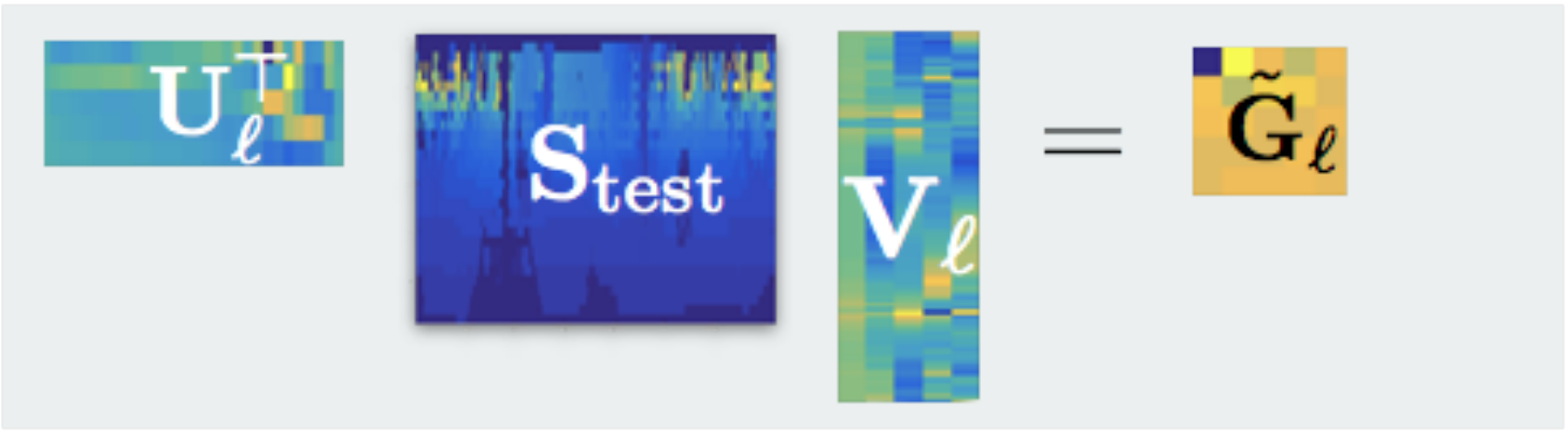} \\\vspace{-35pt}
    \hspace{150pt}The ``signature''\\\vspace{15pt}
  \caption{\small Localizing based on a scan. Each test scan, after matricization (as described in Section~\ref{sec:local}), is processed by each $\mathbf{U}_\ell$ and $\mathbf{V}_\ell$ to form ``signatures'' $\tilde{\mathbf{G}}_\ell$, which are then compared (in Frobenius norm sense) to the core tensors $\underline{\mathbf{G}}_\ell$ of TensorMap for best match.}
  \label{fig:detect} 
  \vspace{-20pt}
\end{figure}

\vspace{-12pt}
\sr{\subsection{ Summary of Our Technique}
\vspace{-4pt}
 In this work, we present a tensor decompositions-based technique for building topological maps using Lidar data. To this end, we first represent the 3D-point cloud Lidar scans as a 3-way tensor.  Next, we learn orthogonal Tucker3 models on partitions of this tensor by exploiting the approximate low multi-linear rank structure, arising from the fact that scans in a local neighborhood -- specifically straight paths -- are similar; see Fig.~\ref{figure:Prep}. Further, we develop a technique to localize in this map by leveraging the ``all-orthogonal'' property of the aforementioned tensor decomposition; see Fig.~\ref{fig:detect}. To the best of our knowledge, this is the first application to exploit the orthogonality of the core tensor slices.}

\vspace{-12pt}
\subsection{Our Contributions}
\vspace{-6pt}
\sr{We make the following contributions: 1) we develop TensorMap\footnote{Details about the implementation can be found at  \url{https://github.com/srambhatla/TensorMap}.}: a technique to build  Lidar-based topological maps using tensor decompositions and perform localization in them, 2) we analyze the efficiency of the proposed representation in terms of its space complexity in comparison to using the full Lidar data, 3) we show the performance of TensorMap for a localization task on real Lidar data, and 4) we demonstrate the robustness properties of the proposed technique to different types of simulated noise (Gaussian and translational).}

The rest of the paper is organized as follows. We formulate the problem and describe TensorMap in Section \ref{probForm}. In Section~\ref{sims}, we discuss parameter selection, simulations results, and other applications, and provide a few concluding remarks in Section \ref{conclusion}.

  \vspace{-8pt}
\section{Problem formulation}
\label{probForm}
  \vspace{-6pt}
We illustrate TensorMap using the Ford campus vision and Lidar dataset \cite{Pandey11}, henceforth referred to as ``the Ford Dataset.'' 
The Ford Dataset contains a set of 3800 Lidar scans corresponding to a loop in downtown Dearborn, Michigan. The trajectory of the scans collected by the Ford Dataset is shown in Fig.~\ref{fig:map}(a). The data is collected using a Velodyne 3D-Lidar scanner which has a vertical field of view (FOV) of $26.3^{\circ}$ (apx. from $-25^{\circ}$ to $4^{\circ}$) and a lateral FOV of $360^\circ$ (from $[-180^{\circ}, 180^{\circ}]$), with the Lidar spinning at 10 Hz. 
\begin{figure*}[th]
\centering
\begin{tabular}{cccc}
\multicolumn{4}{c}{ ~~~~\epsfig{file=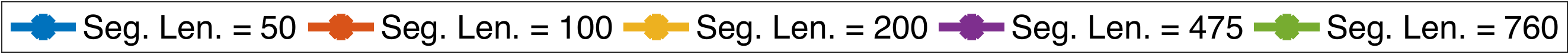,width=0.5\linewidth,clip=} }\\
    \epsfig{file=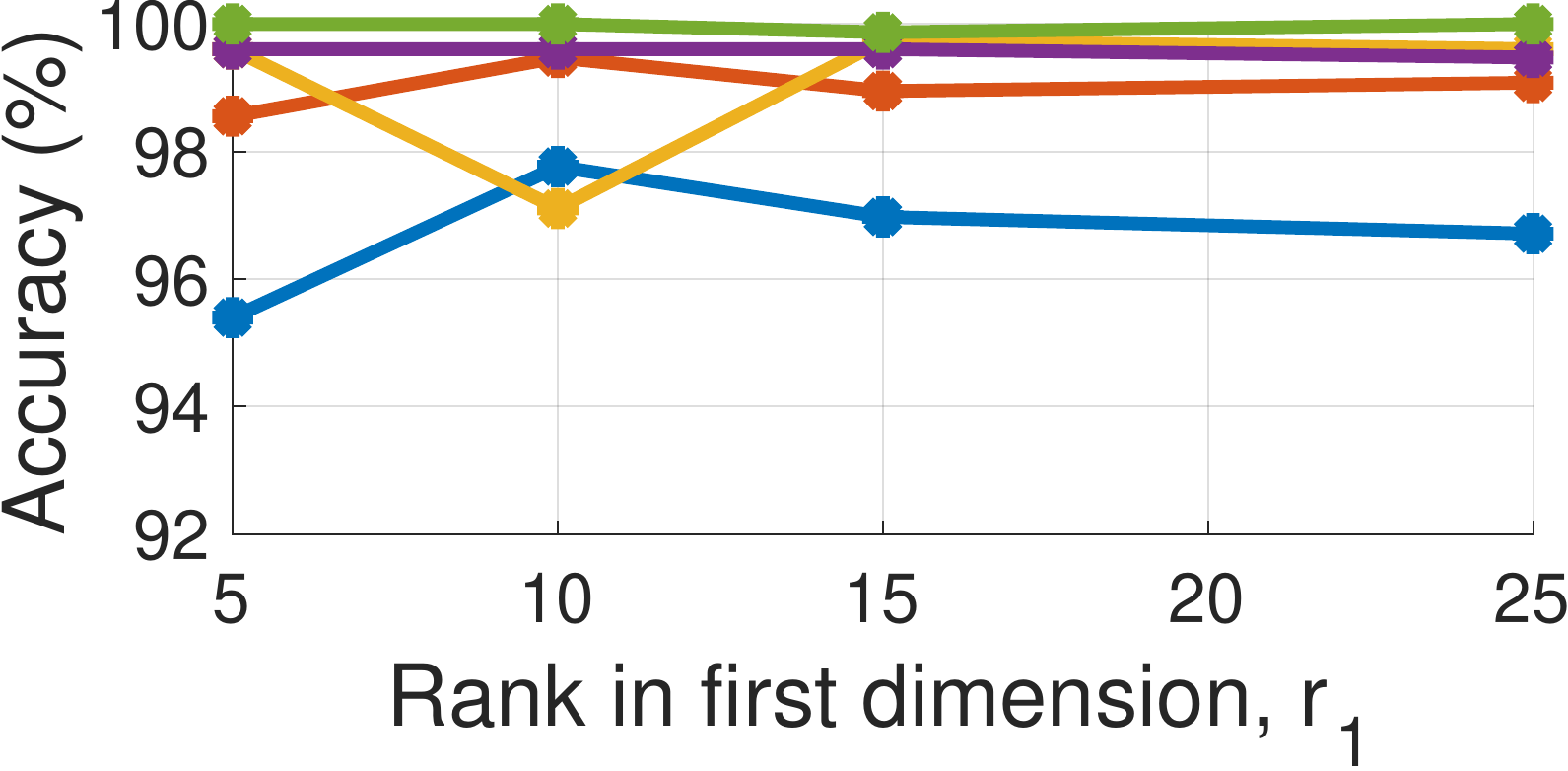,width=0.22\linewidth,clip=} & \epsfig{file=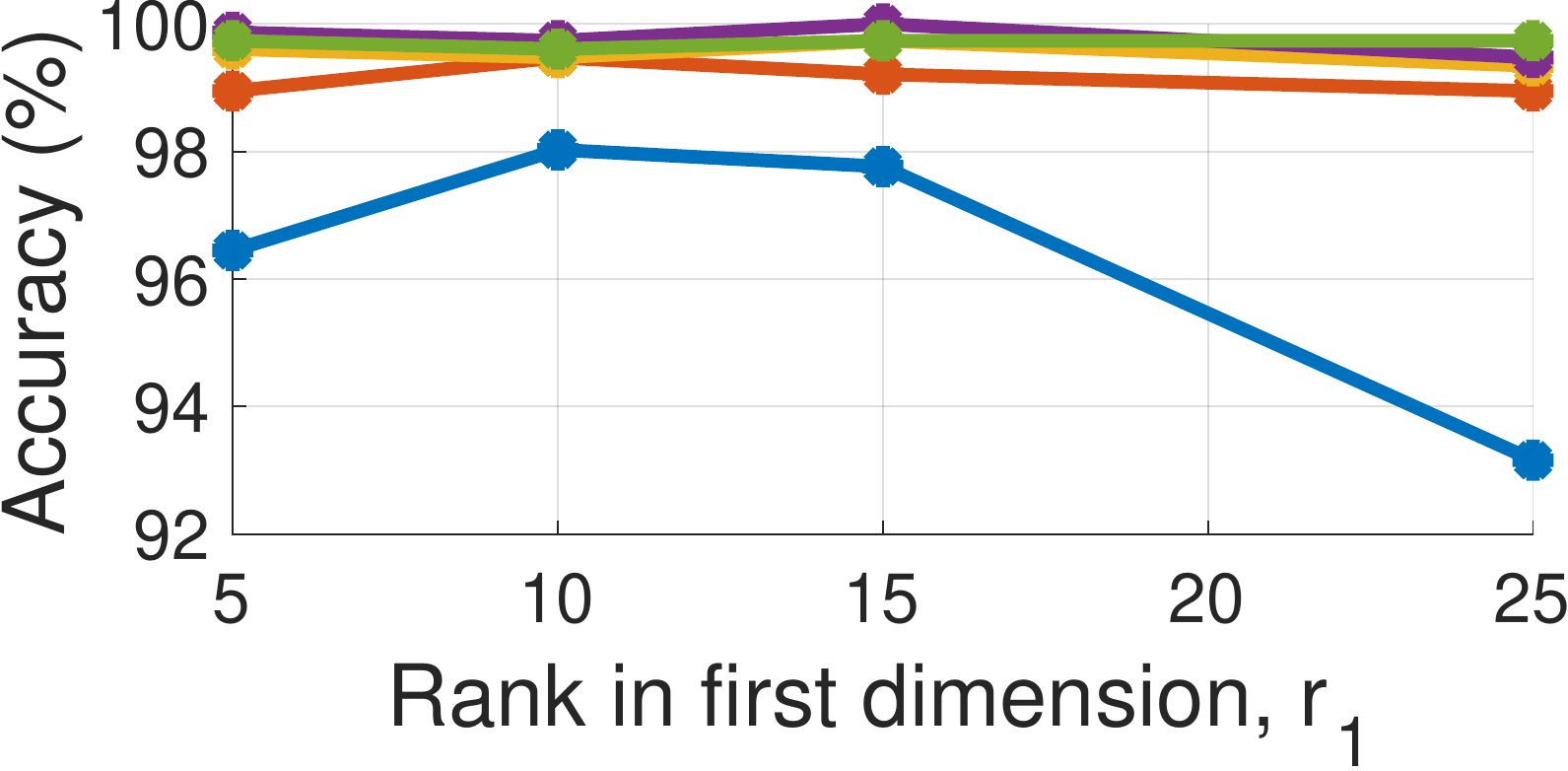,width=0.22\linewidth,clip=} &
    \epsfig{file=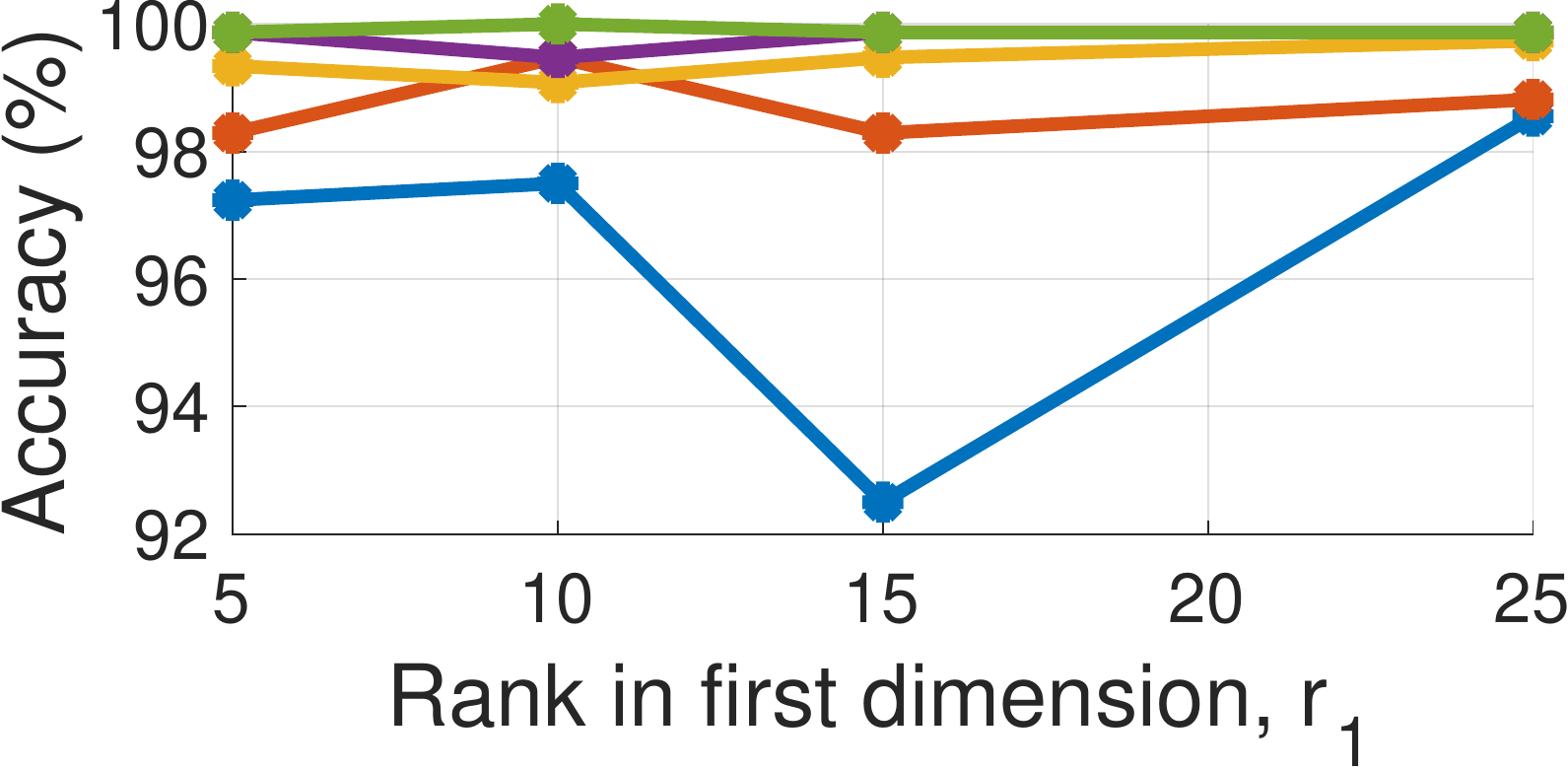,width=0.22\linewidth,clip=} &
    \epsfig{file=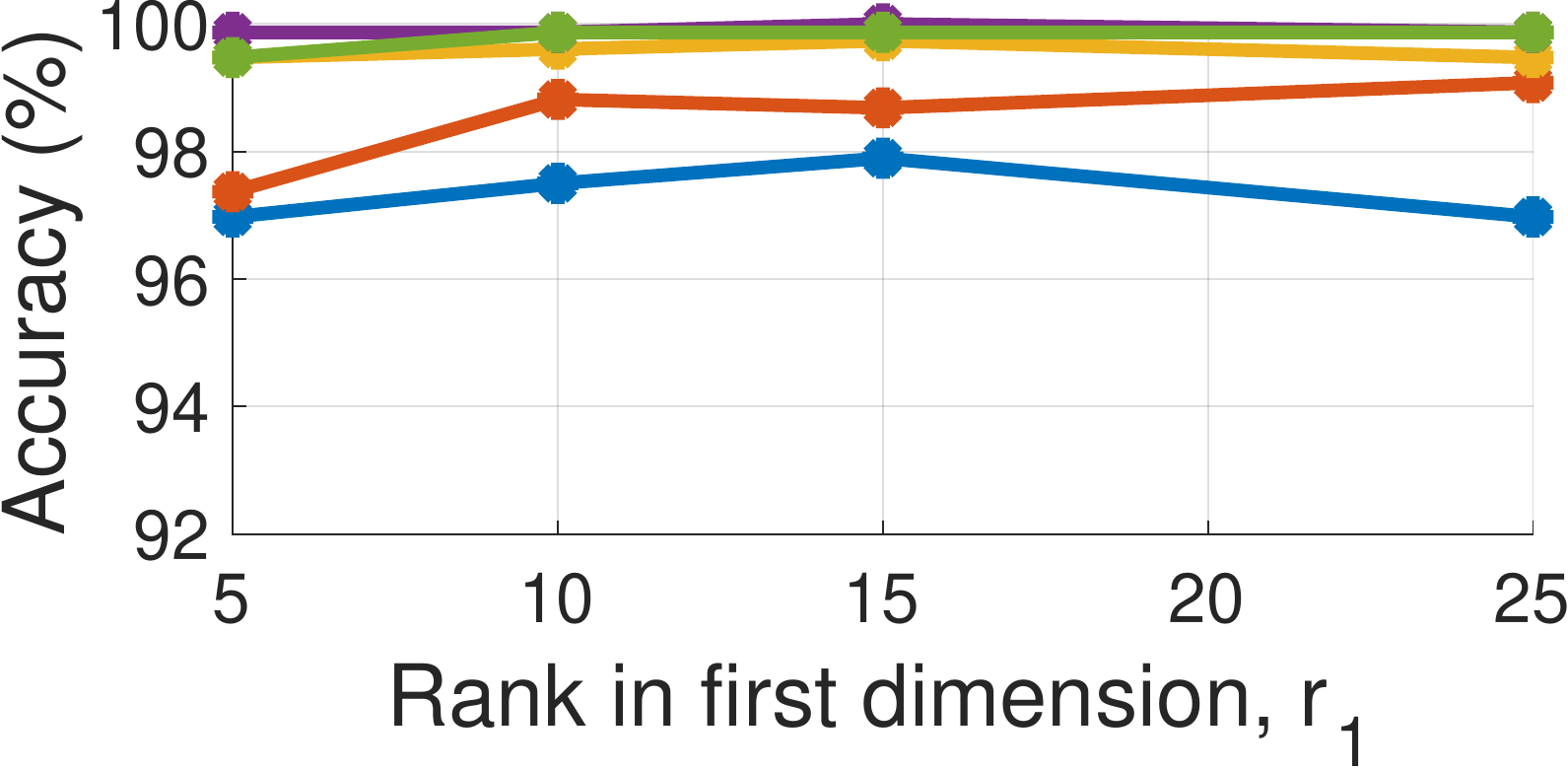,width=0.22\linewidth,clip=} \vspace{-5pt}\\
    (a) $r_2 = 5$& (b) $r_2 = 10$& (c) $r_2 = 15$& (d) $r_2 = 25$\\ 
 \epsfig{file=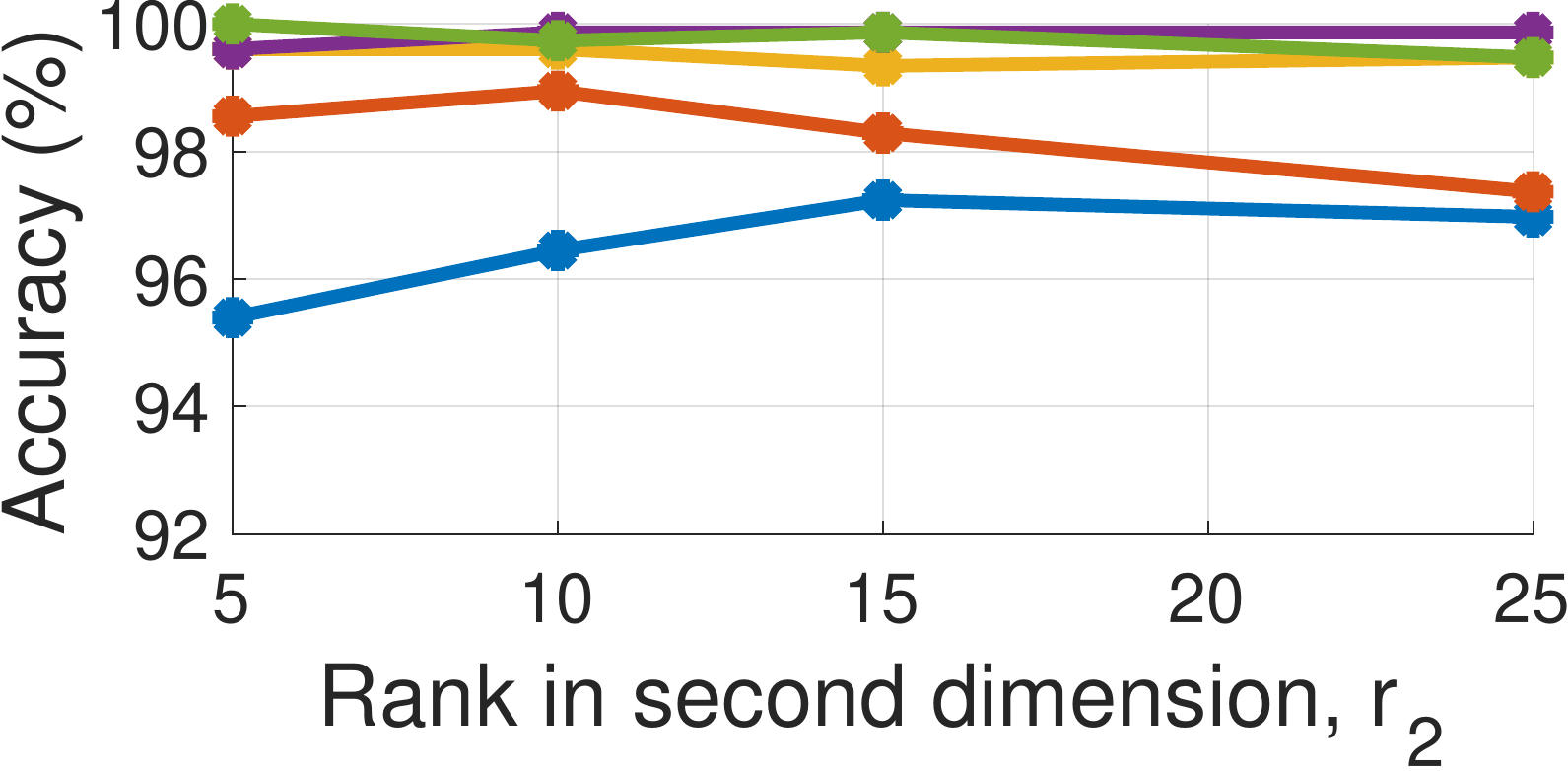,width=0.22\linewidth,clip=} & \epsfig{file=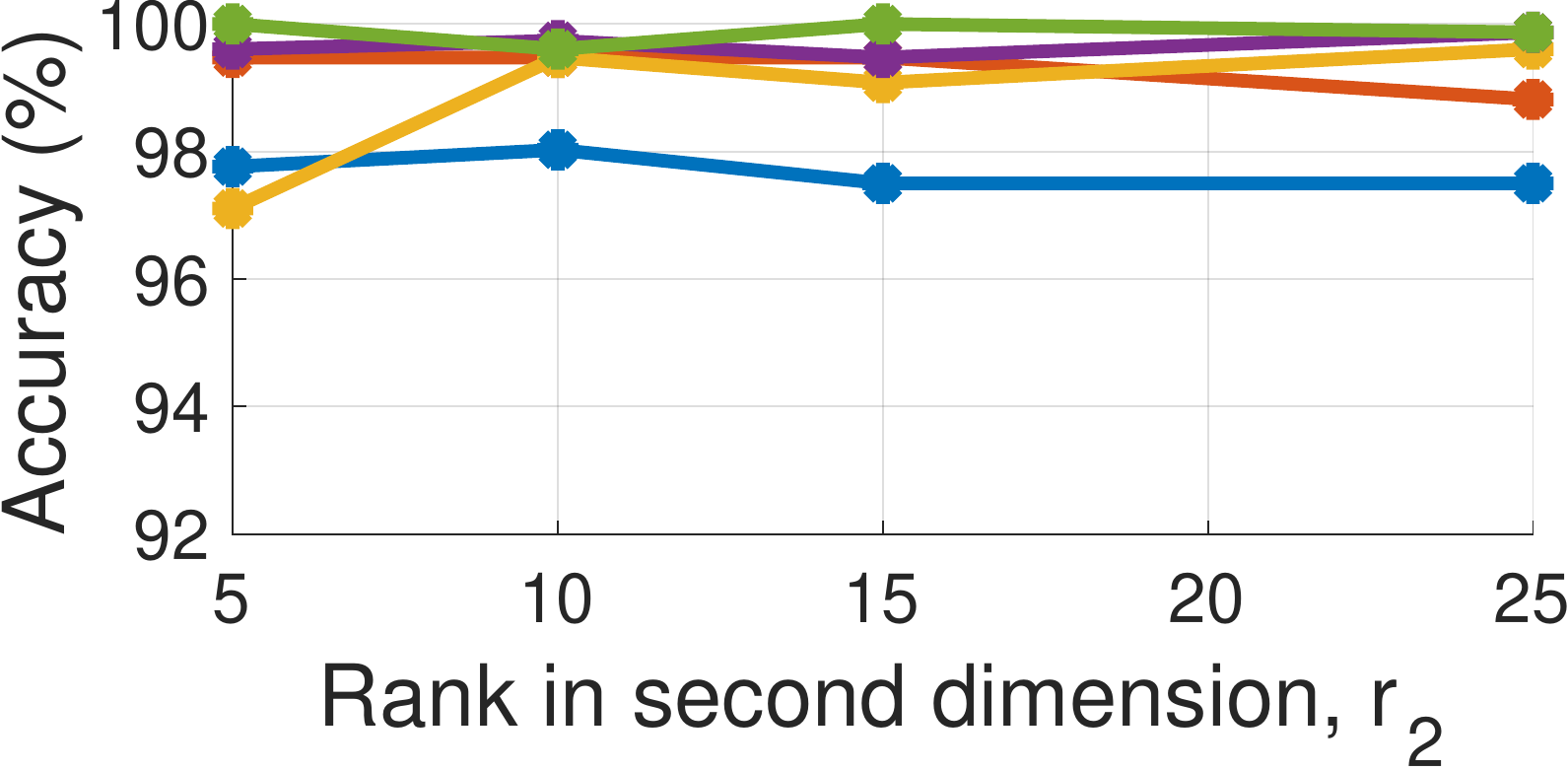,width=0.22\linewidth,clip=} &
    \epsfig{file=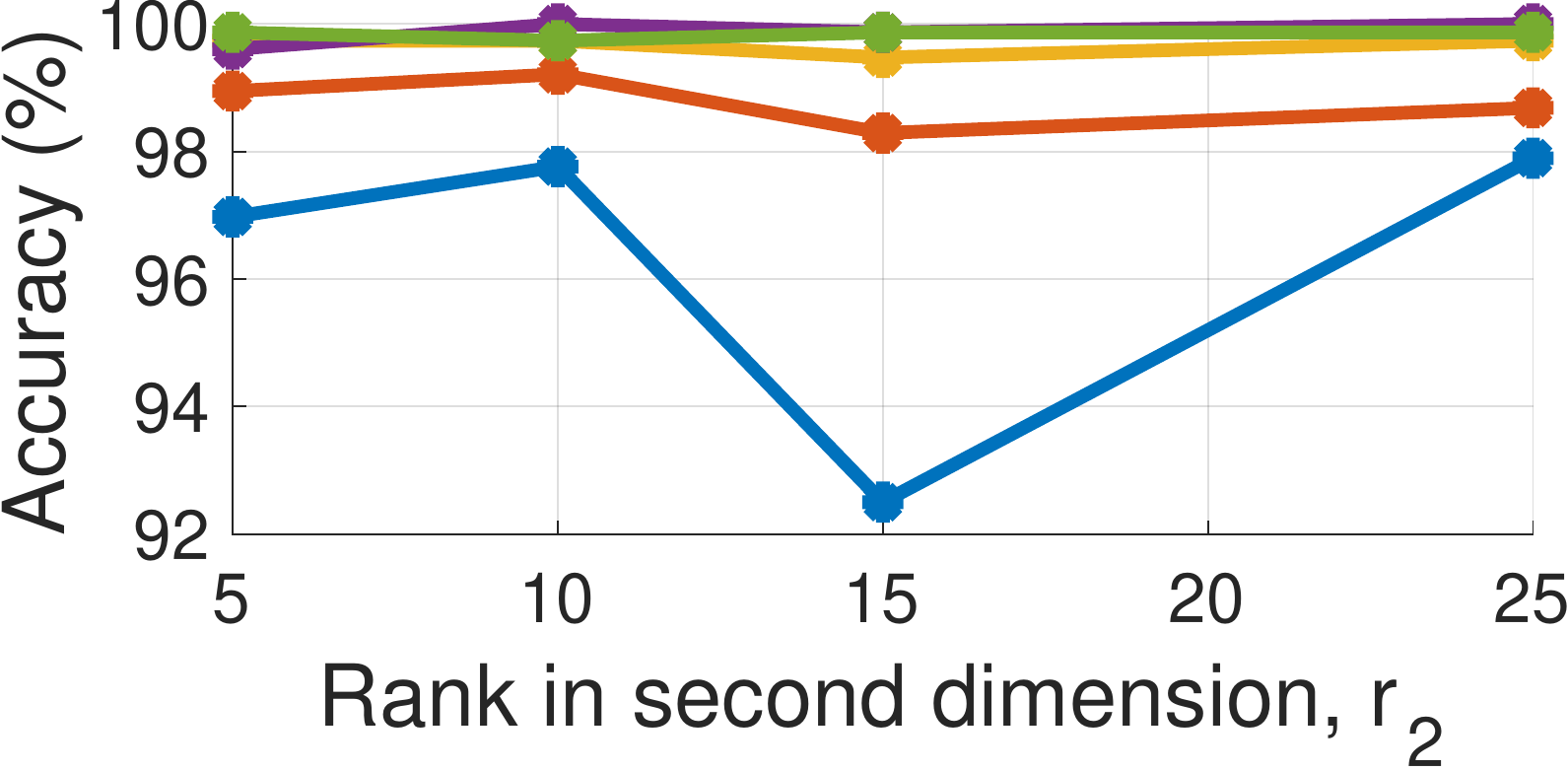,width=0.22\linewidth,clip=} &
    \epsfig{file=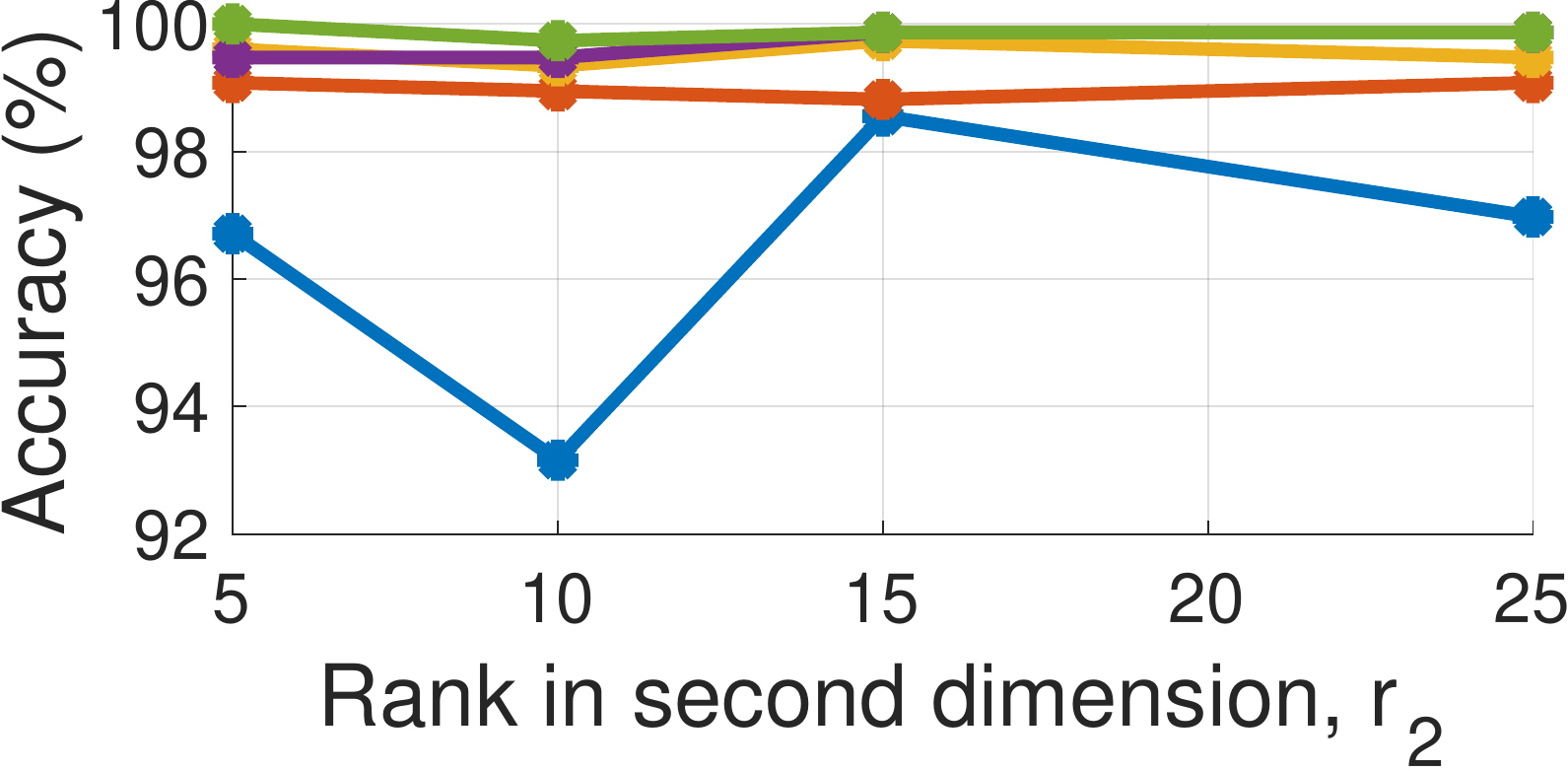,width=0.22\linewidth,clip=} \vspace{-5pt}\\
    (e) $r_1 = 5$& (f) $r_1 = 10$& (g) $r_1 = 15$&(h) $r_1 = 25$\\
 \vspace{-25pt}
\end{tabular}
\begin{tabular}{ccccc}
\epsfig{file=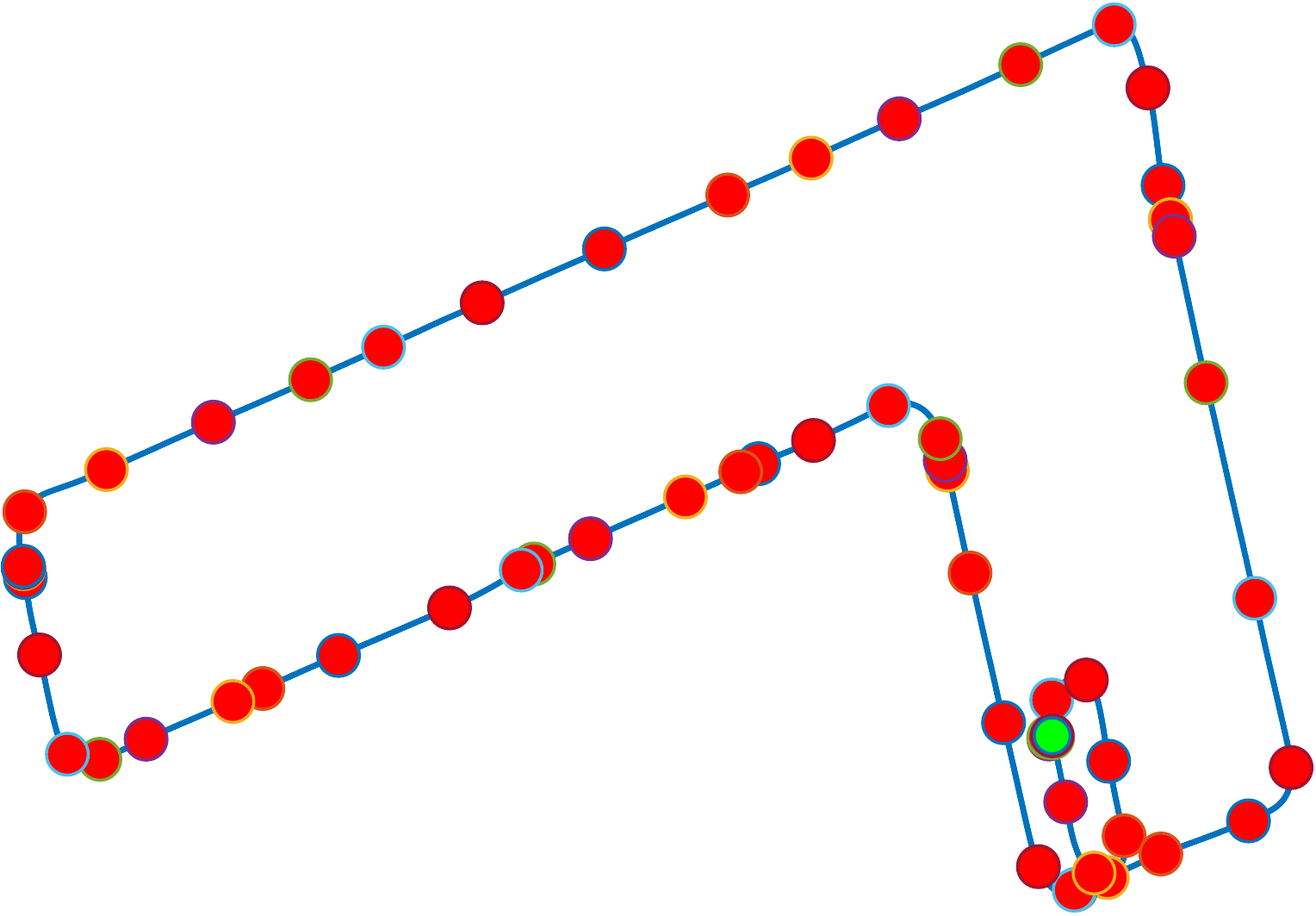,width=0.1\linewidth,angle=337,clip=} &
\epsfig{file=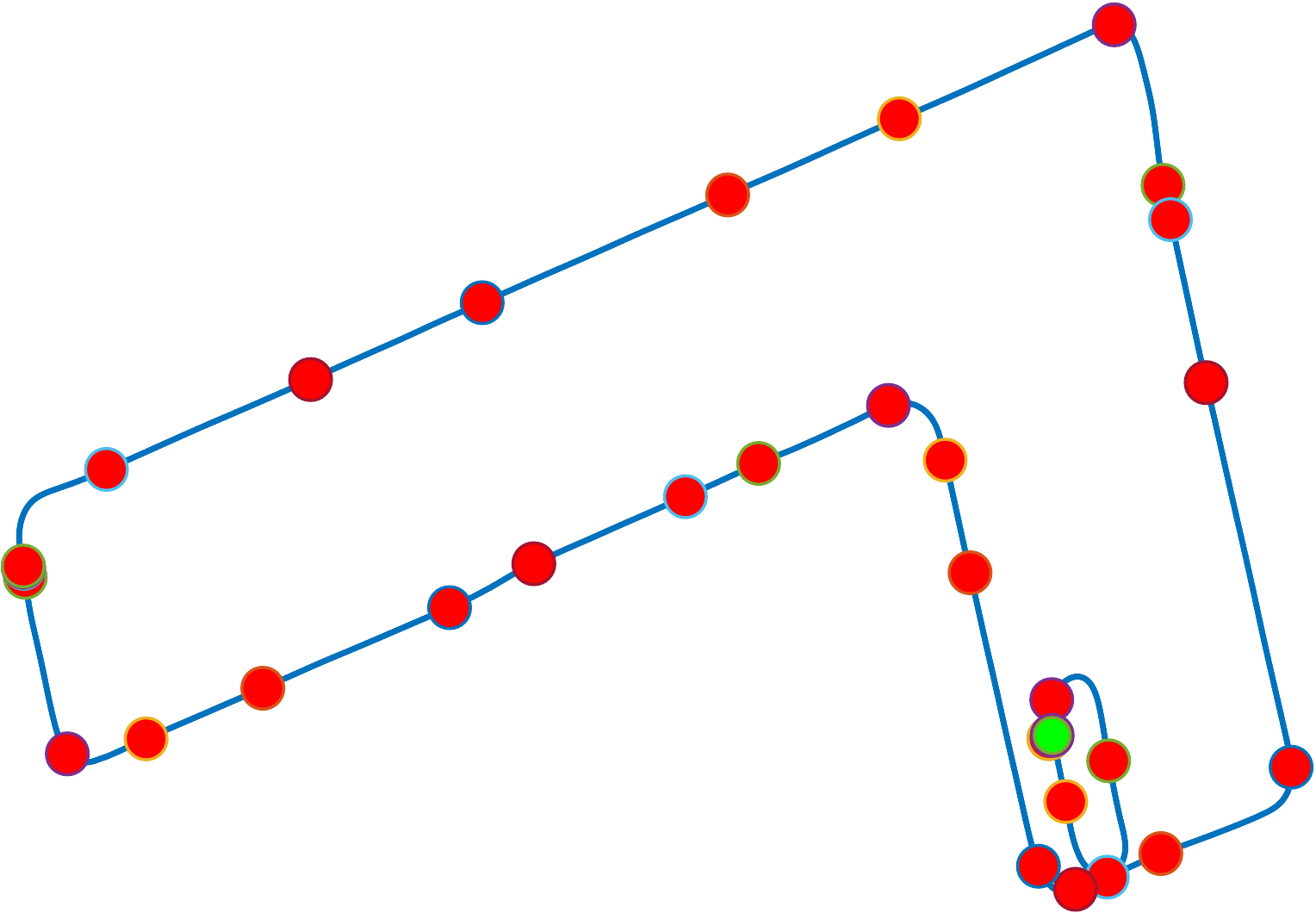,width=0.1\linewidth,angle=337,clip=} &
\epsfig{file=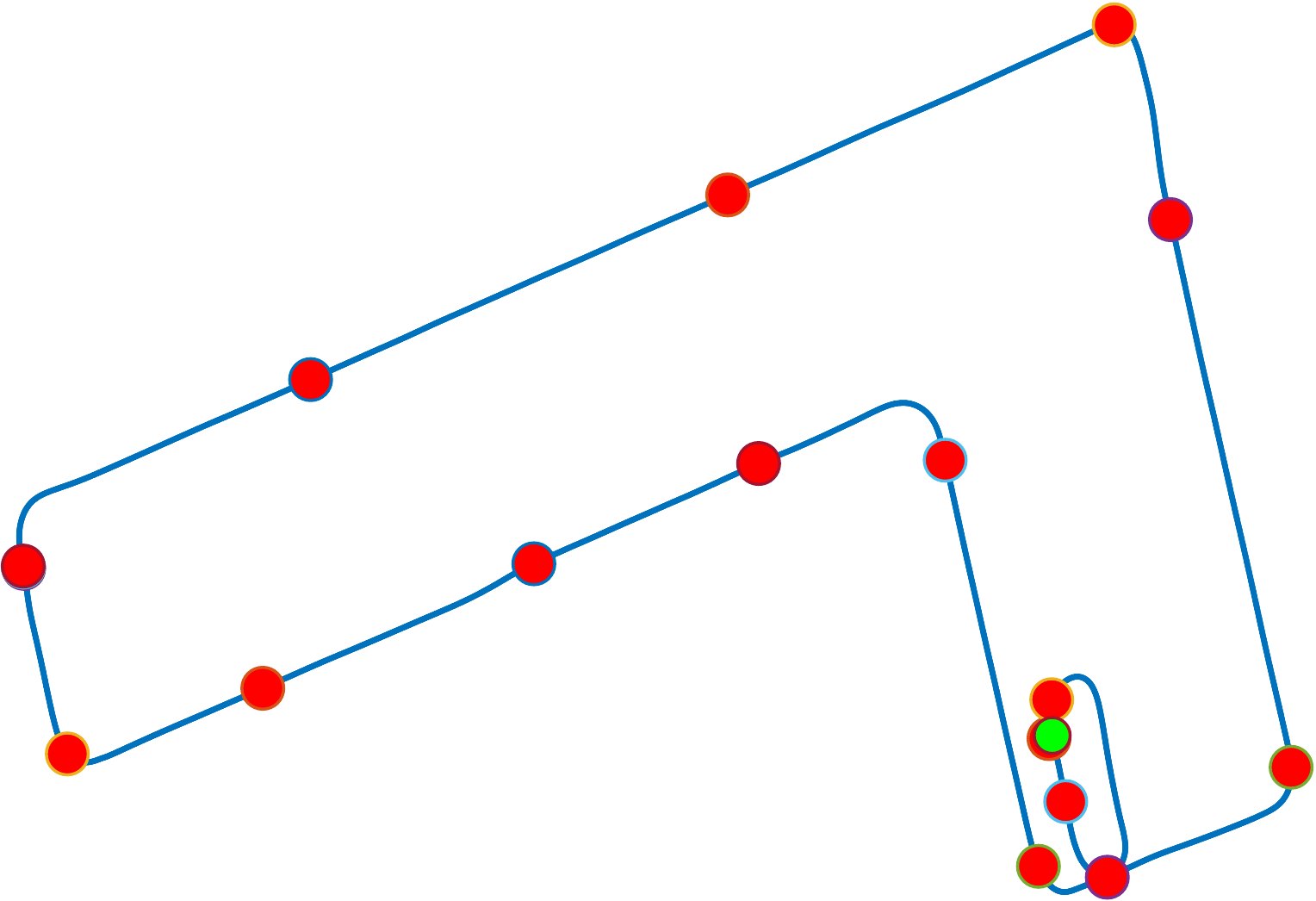,width=0.1\linewidth,angle=337,clip=} &
\epsfig{file=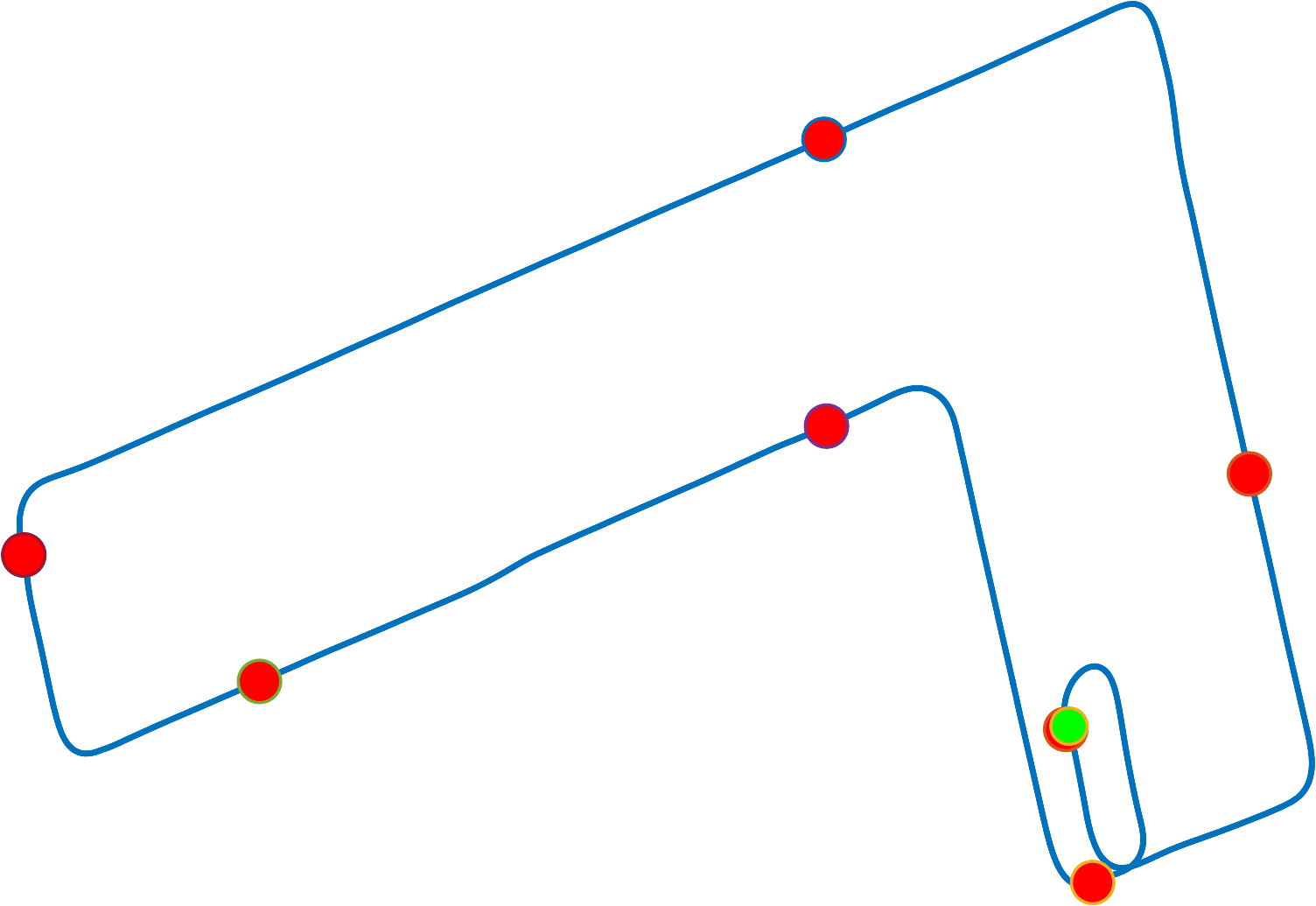,width=0.1\linewidth,angle=337,clip=} &
\epsfig{file=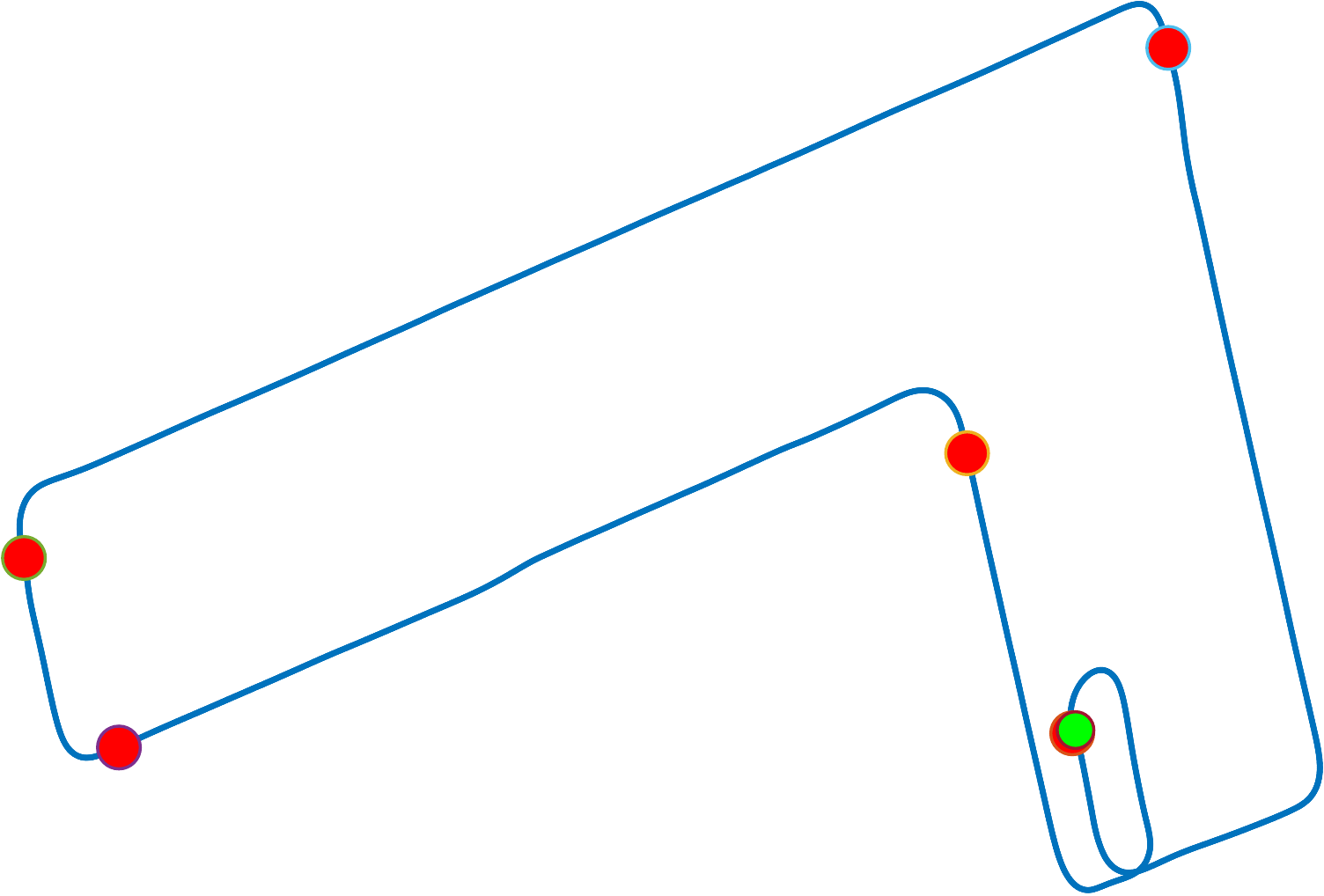,width=0.1\linewidth,angle=337,clip=}\\
(i) Seg. Len. $k=50$&
(j) Seg. Len. $k=100$&
(k) Seg. Len. $k=200$&
(l) Seg. Len. $k=475$&
(m) Seg. Len. $k=760$
\end{tabular}
 \vspace{-8pt}
\caption{\small Effect of choice of \{$r_1, r_2, k$\} on the performance accuracy. Panels (a-d) show the effect of choice of segment lengths $k$ and varying $r_1$ for fixed $r_2 = 5, 10, 15, ~\text{and} ~25$, respectively. Similarly,  panels (e-h) show the effect of choice of segment lengths $k$and $r_2$ for fixed $r_1= 5, 10, 15, ~\text{and} ~25$, respectively. Here, segment lengths $k$ considered are $50, 100, 200, 475, ~\text{and}~ 760$. Panels (i)-(m) show the nodes for each segment corresponding to choice of $k$ (in red), with the start/end point of the path denoted in green.  }
\label{figure:Results}
\vspace{-18pt}
\end{figure*}
\vspace{-8pt}
\subsection{Modeling Lidar data as a Tensor} \label{sec:model_data}
\vspace{-4pt}
Each scan in the dataset is a list of about 77,000 \textit{returns} or a point cloud represented in 3D Cartesian coordinates i.e. $(x, y, z)$ corresponding to the position of objects reflecting the incident laser, as shown in Fig.~\ref{figure:Prep} (a). Here, the number of returns per scan depends on the scene. To represent Lidar scans as a tensor, we first convert the the data to polar coordinates, which results in a list of returns expressed as $(\rho, \theta, \phi)$, where $\rho$ is the range, $\theta$ is the elevation and $\phi$ is the azimuth. Next, we form a matrix with rows corresponding to elevation angles $\theta$, and columns corresponding to azimuth angles $\phi$, by rounding these to whole angles (this discretization is a design choice). Then, for each entry in the list of returns in polar coordinates, we place the range values ($\rho$) at the rounded-off ($\theta,\phi$) location, as shown in Fig.~\ref{figure:Prep} (b). Due to this quantization (of the $\theta$ and $\phi$), multiple returns may get mapped to a single entry of the matrix. For the given sensor, $\theta$ is restricted between $[{-25}^\circ, {4}^\circ]$ and $\phi$ between $[-180^{\circ}, 180^{\circ}]$. Therefore, each scan is transformed to a $30 \times 361$ matrix, and collecting all the scans, results in a tensor $\underline{\mathbf{X}}$.  
Therefore, for the Ford data set $\underline{\mathbf{X}} \in 30\times 361 \times 3800$.

\vspace{-10pt}
\subsection{Building TensorMap}
\vspace{-6pt}
 For learning the topological map, we use the orthogonal Tucker decomposition to exploit the low mode-rank (in two of the three modes) structure of the tensor. Lidar data is particularly amenable to this model because the scene at each step is highly correlated to the previous one. 
To leverage this relationship, let  $\underline{\mathbf{X}}$ denote a tensor in $ \mathbb{R}^{\rm I\times J \times K}$ containing all scans corresponding to a map. Next, let $\underline{\mathbf{X}}_\ell \in  \mathbb{R}^{\rm I\times J \times k}$ denote length-$k$ disjoint partitions of $\underline{\mathbf{X}}$ for each $\ell = \{1, 2, \dots, \rm L\}$ for  ${\rm L = K}/k$, where we assume that $k$ divides $K$ perfectly; see Fig.~\ref{figure:Prep}(c). 
As a result, we have short tensors  $\underline{\mathbf{X}}_\ell$ for each length-$k$ segment along the path whose orthogonal Tucker3 decomposition can be written as 
   \vspace{-4pt}
\begin{equation*}
   \vspace{-4pt}
{\rm vec}(\underline{\mathbf{X}}_\ell) = (\mathbf{U}_\ell \otimes \mathbf{V}_\ell \otimes \mathbf{W}_\ell) \bar{\mathbf{g}}_\ell.
\end{equation*}
Here,  ``$\otimes$'' denotes kronecker product, $\mathbf{U}_\ell \in \mathbb{R}^{\rm I\times r_1}$, $\mathbf{V}_\ell \in \mathbb{R}^{\rm J\times r_2}$ and $\mathbf{W}_\ell \in \mathbb{R}^{\rm k \times k}$ denote the factors where $r_1 \le\rm I$ and $r_2 \le \rm J$, and $\bar{\mathbf{g}}_\ell$ denotes the vectorized core tensor $\underline{\mathbf{G}} _\ell$ shown in Fig.~\ref{figure:Prep}(c). 
Note that, to preserve the position information we do not compress along the third dimension of the segment tensor $\underline{\mathbf{X}_\ell}$, i.e., we set $\mathbf{W}_\ell = \mathbf{I}$, where $\mathbf{I}$ denotes an $\rm k \times k$ identity matrix.   
The core tensor $\underline{\mathbf{G}} _\ell \in \mathbb{R}^{\rm r_1\times r_2 \times k}$ along with factors $\mathbf{U}_\ell$ and $\mathbf{V}_\ell$ corresponding to each segment form the TensorMap, as shown in Fig.~\ref{figure:Prep}(d).

\vspace{-10pt}
 \subsection{Localizing in TensorMap}\label{sec:local}
 \vspace{-6pt}
Since each ${\rm r_1\times r_2}$ slice of the core tensor $\underline{\mathbf{G}} _\ell \in \mathbb{R}^{\rm r_1\times r_2 \times k}$ (corresponding to the scans in a segment) is orthogonal to the other slices, each slice of the core tensor can be viewed as a ``signature'' of the associated scan.
 As shown in Fig.~\ref{fig:detect}, we exploit this property for localization. Specifically, to localize any test scan (point cloud), we first convert it into a matrix $\mathbf{S}_{\text{test}}$ as described in Section~\ref{sec:model_data}. Next, we form ``signature" $\tilde{\mathbf{G}}_\ell$ corresponding to $\mathbf{S}_{\text{test}}$  as
    \vspace{-3pt}
\begin{equation*}
   \vspace{-3pt}
\tilde{\mathbf{G}}_\ell= \mathbf{U}_\ell^\top \mathbf{S}_{\text{test}} \mathbf{V}_\ell, 
\end{equation*}
for all $\ell \in \{1, 2, \dots, \rm L\}$. Then, we find the closest matching core tensor slice $\mathbf{G}_\ell$ (in Frobenius norm sense) across all segments. This process identifies the scan that is a closest match to the test scan, hence also identifies the segment.

\vspace{-10pt}
\subsection{Memory Considerations}
\vspace{-6pt}
\sr{We consider the space complexity of TensorMap for its implementation on real-world systems and embedded platforms.} We propose to learn a orthogonal Tucker3 model for each length-$k$ segment, and there are $L$ such models to be learnt. Therefore, the total number of memory units required to store TensorMap are,
\vspace{-5pt}
\[\rm L(Ir_1 + Jr_2) + Kr_1 r_2.\vspace{-5pt}\]
 This storage requirement is significantly smaller than the original tensor, i.e. $\rm IJK$, for small values of $\rm r_1,~r_2$ and $\rm L$. Note that we do not store $\mathbf{W}_\ell$ since in each case it is an identity matrix.

Interestingly, the expression above supports longer segments which still yield a lower error for smaller $\rm r_1$ and $\rm r_2$. In the context of maps, this means that scans of a segment should be accumulated as long as they are similar to each other. 
Therefore, suitable segment length is closely related to the number of straight line paths in the map.  Note that, although we consider a fixed segment length for the current exposition, there is no requirement that the segments be of equal length. We leave exploration of these extensions to future work.

  \vspace{-10pt}
\section{Numerical Evaluations}\label{sims}
  \vspace{-6pt}
\begin{figure*}[th]
\centering\setlength{\belowcaptionskip}{0pt}
\begin{tabular}{ccc}
    \epsfig{file=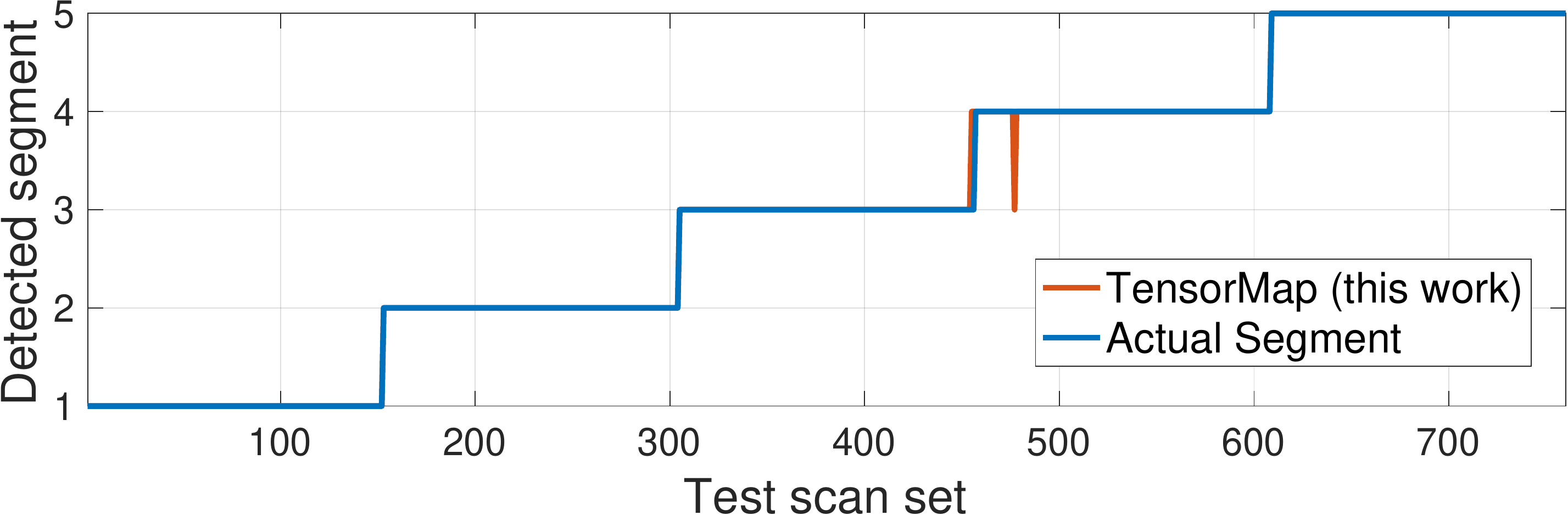,width=0.36\linewidth,clip=} & \epsfig{file=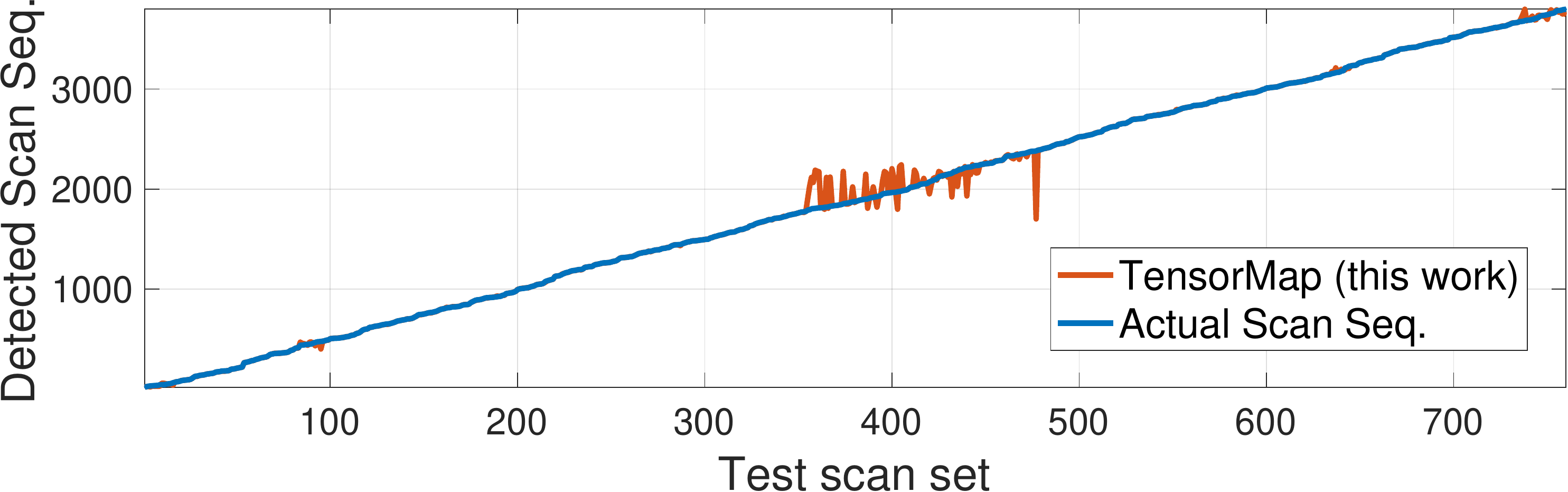,width=0.37\linewidth,clip=} &
    \epsfig{file=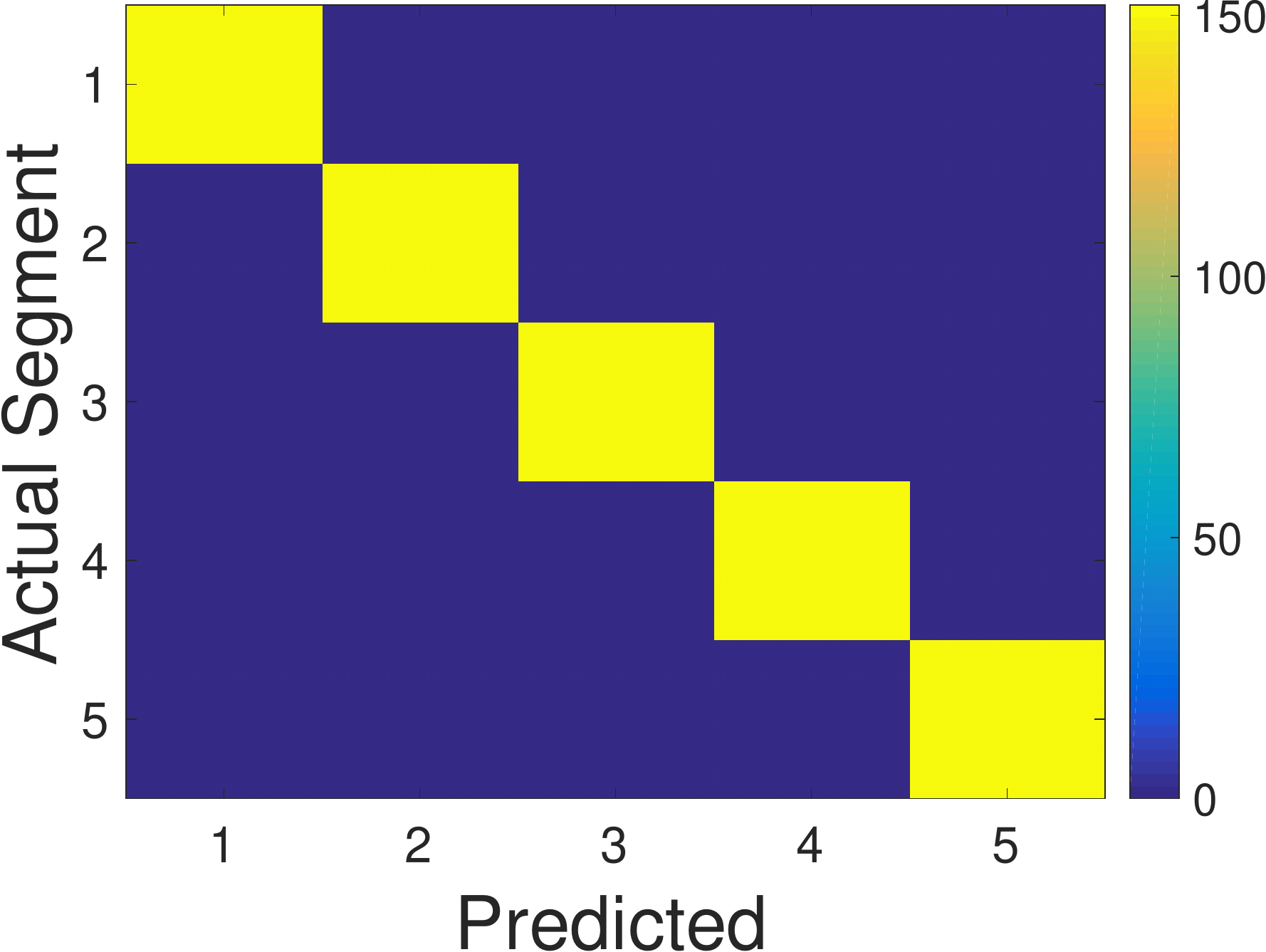,width=0.16\linewidth,clip=} \vspace{-3pt}\\
    (a) & (d) & (g)\\
    \epsfig{file=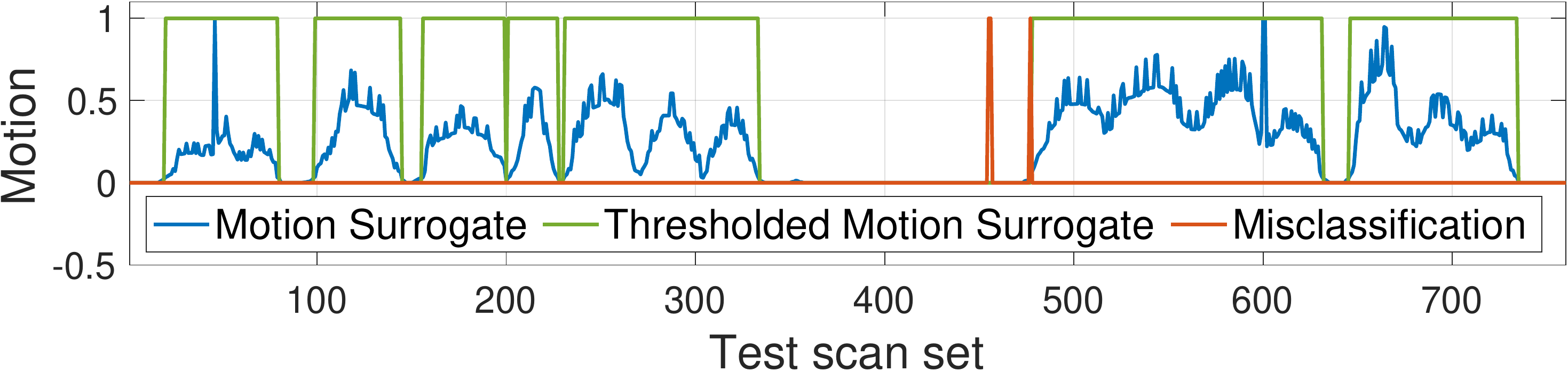,width=0.36\linewidth,clip=} & \epsfig{file=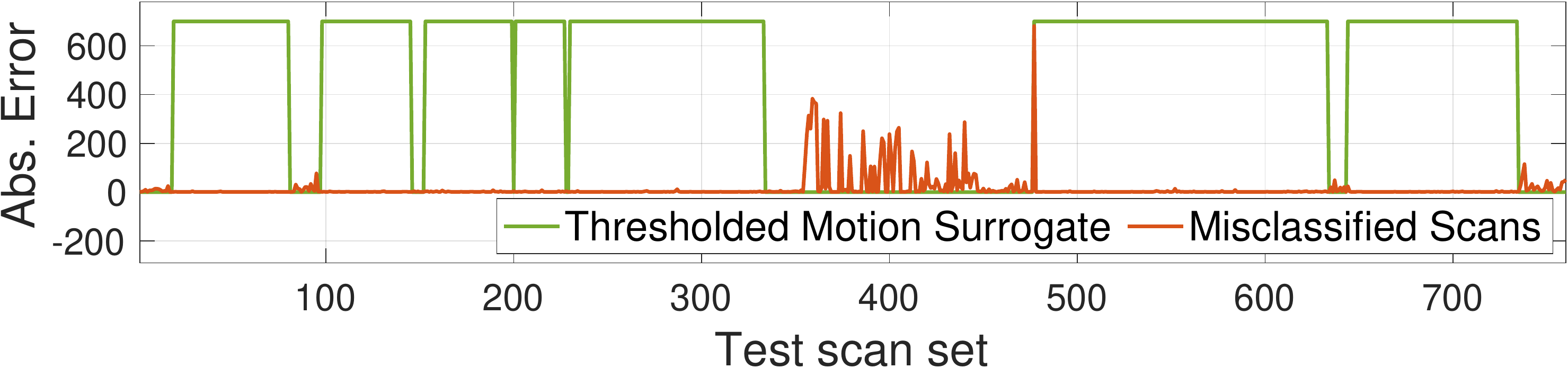,width=0.37\linewidth,clip=} &
    \epsfig{file=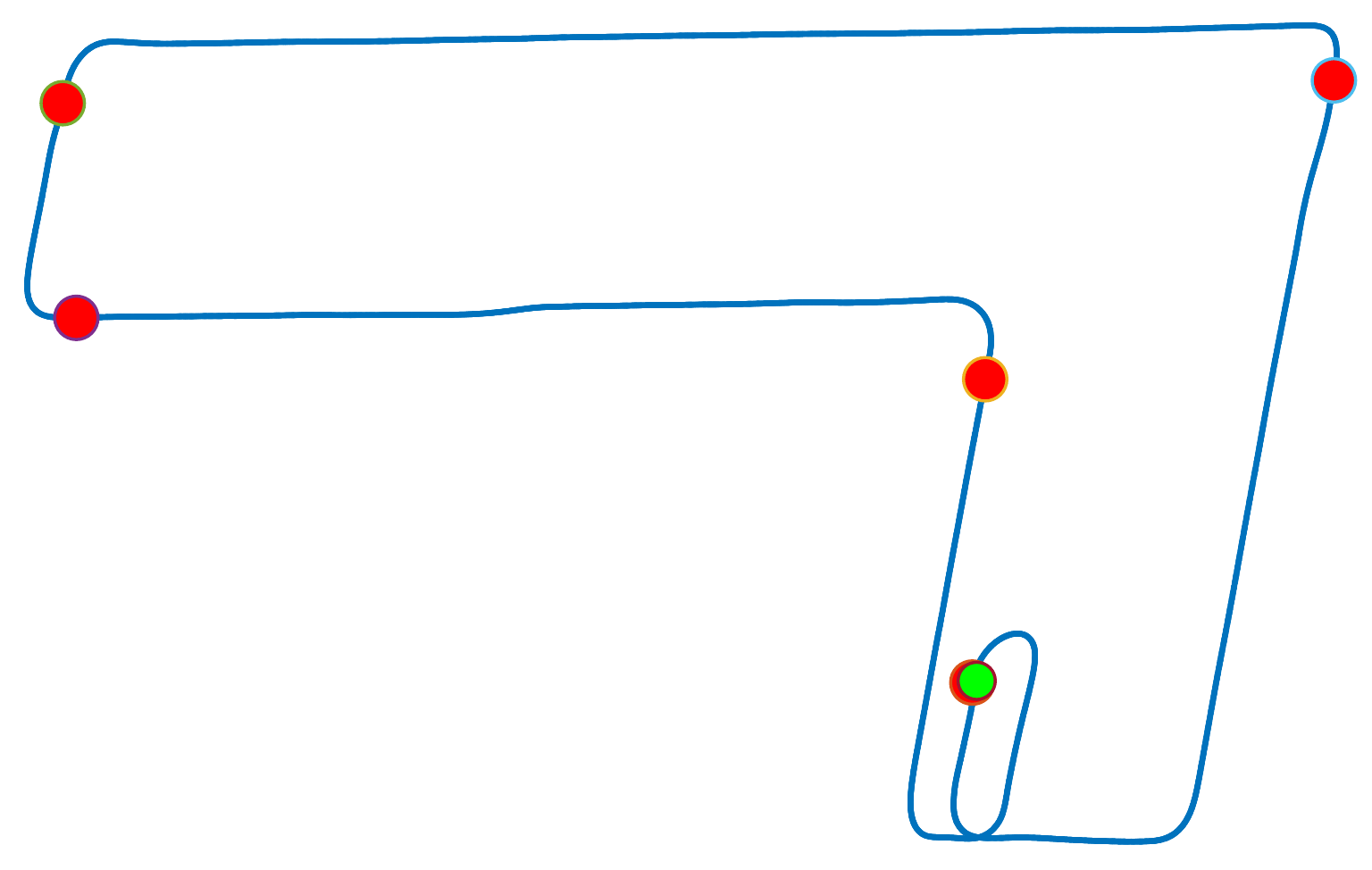,width=0.12\linewidth,clip=}\vspace{-3pt}\\
    (b) & (e) &(h)\\
    \epsfig{file= 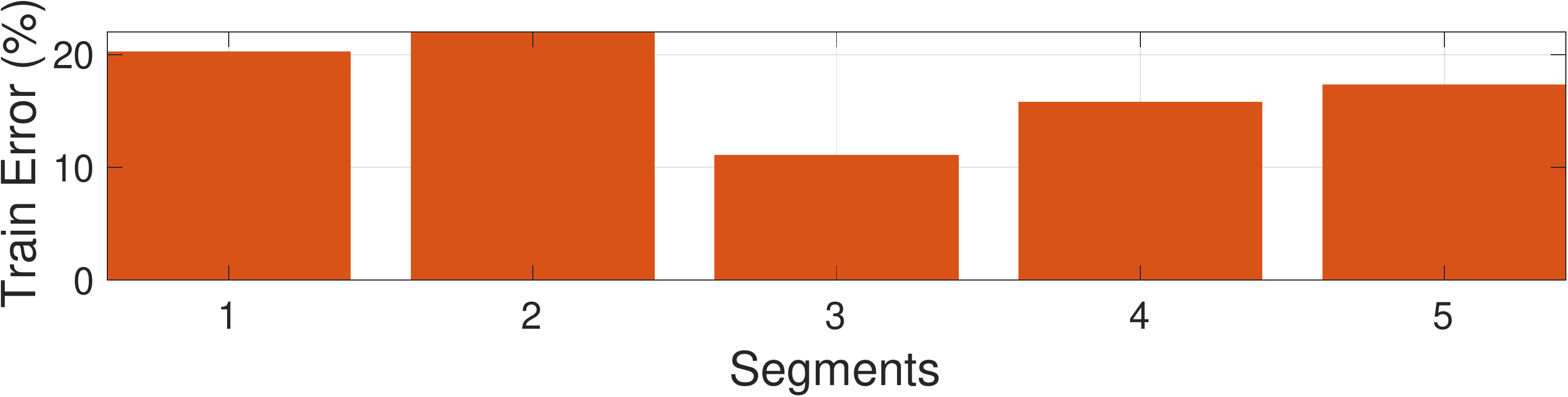,width=0.35\linewidth,clip=} & \epsfig{file=tucker_err_gsip-crop.pdf,width=0.35\linewidth,clip=} &\vspace{-3pt}\\
    (c) & (f) & 
    \vspace{-5pt}
\end{tabular}
\caption{\small Performance of TensorMap on the Ford Dataset with $\{r_1, r_2, k\}$ chosen as $\{5, 5, 760\}$, respectively. Panel (a) shows the classification of test scans into segments. The corresponding surrogate for velocity (blue), the decision of vehicle movement (green), and the errors made by TensorMap (red) are shown in panel (b). Notice how majority of the errors occur when the vehicle is stationary. Panels (c) and (f) show the relative error between the original segment tensor and the model learnt by TensorMap. Panel (d)  shows the scan classification performance of the technique, actual test set (blue) the closest (Frobenius norm) train set scan found by TensorMap. The corresponding decision of vehicle movement (green) and the errors made (red) are shown in (e). Panel (g) shows the confusion matrix corresponding to the classification of test scans to segments shown in (a), and (h) shows the nodes of TensorMap (red) superimposed on the actual map (blue).}
\label{figure:Results_1}
  \vspace{-18pt}
\end{figure*}

We now discuss the performance of the proposed procedure on the Ford Dataset. 
\vspace{-10pt}
\subsection{Experimental Set-up}
\vspace{-6pt}
We evaluate the performance of TensorMap based on its classification accuracy of assigning test scans to their respective segments, using a $80:20$ - Train : Test split of scans in each segment. To this end, we first learn orthogonal Tucker representations (TensorMap) on the training data for each segment using the HO-SVD algorithm \cite{De2000}\cite{Kolda09}. We also analyze the within-segment classification performance by analyzing the train scan sequence which was found closest to the test sequence. 

\vspace{-10pt}
\subsection{Selecting the Parameters}
\vspace{-4pt}
 There are a few design parameters that we need to choose, namely the length of the segment $k$, and the number of columns $r_1$ and $r_2$ in factors $\mathbf{U}_\ell$ and $\mathbf{V}_\ell$, respectively. To find the best choice(s), we search over various values of $r_1$, $r_2$, and $k$, to arrive at a $\{r_1, r_2, k\}$ which yields highest accuracy, while being efficient in terms of the storage requirements. 
 
 Fig.~\ref{figure:Results} shows accuracies over different choices of $\{r_1, r_2\}$, and segment lengths $k$. We observe that for a specific choice of $r_1$ and $r_2$, the segment classification performance is better for longer segments as compared to shorter ones. This is because scans in shorter segments are very similar to those in neighboring segments; see Fig.~\ref{figure:Results} (i)-(m). Also, although longer segments choices sometimes perform better for larger values of $r_1$ and $r_2$, we prefer smaller $r_1$ and $r_2$ to reduce the computational and memory overhead. Overall, by this analysis, we arrive at the choice of $\{5, 5, 760\}$ for $\{r_1, r_2, k\}$, respectively. 
 
 \vspace{-10pt}
 \subsection{Results} 
 \vspace{-4pt}
 In Fig.~\ref{figure:Results_1},  we present the results for $\{r_1, r_2, k\}$ chosen  as $\{5, 5, 760\}$, respectively. 
 We observe that our method identifies the test segments accurately, except for two scans; see Fig.~\ref{figure:Results_1}(a). To investigate these misclassifications, we turn to Fig.~\ref{figure:Results_1}(b), which shows the relationship of the errors with the motion surrogate, which is formed by evaluating the norm of change in $6$-DOF pose -- provided by the Ford Dataset -- of the vehicle. We observe that the errors seem to arise only when the vehicle is stationary. \sr{This is due to the fact that the scan acquisition process does not stop when the vehicle is not moving. As a result, scenes in consecutive segments can be very similar to each other. 
 However, attributing scans to any one of the these segments does not adversely effect the localization performance. Therefore, to account for this effect we report errors on parts where the vehicle is moving, using the motion surrogate.}
 
  In panel Fig.~\ref{figure:Results_1}(d) and (e), we show the actual train scan (scan sequence number) found to be the closest to the test set and the misclassified scans, respectively. 
We note that when the vehicle is in motion, TensorMap indeed performs very well. In practice, we can run TensorMap only when the vehicle is in motion, holding the currently estimated value when the vehicle is stopped.  

\sr{We also report the error between the original segment tensor and the orthogonal Tucker3 model learnt in Fig.~\ref{figure:Results_1}(c) and (f), replicated to improve readability. 
 Further, Fig.~\ref{figure:Results_1} panel (g) shows the corresponding confusion matrix for segment classification problem shown in Fig.~\ref{figure:Results_1}(a). Also, the topological map learnt is shown in panel (h). Notice that the nodes of this topological map are not spaced uniformly, this is due to the movement of the vehicle. }
\vspace{-10pt}
\subsection{Effect of Gaussian noise and Translations}
\vspace{-6pt}
We now study the effect of Gaussian noise and translations on the performance of TensorMap. Here, we generate the noisy tensor by adding zero-mean Gaussian random noise of variance $\sigma^2$ to each coordinate of the Lidar scan, and process these noisy Lidar scans using the procedure described in Section~\ref{sec:model_data}.

\begin{figure}[h]
  \centering
  \begin{tabular}{cc}
    \includegraphics[width=0.2\textwidth]{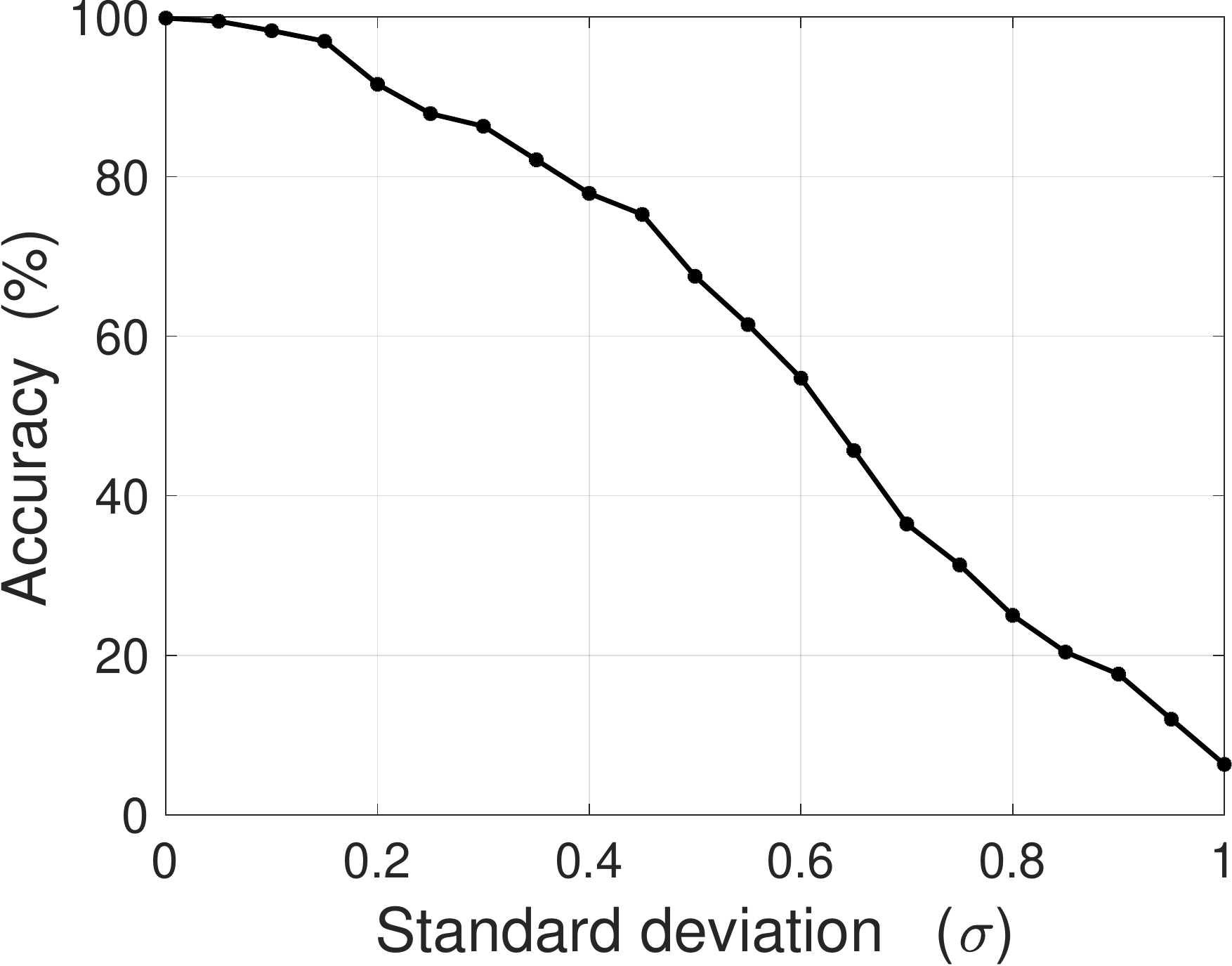} & \includegraphics[width=0.2\textwidth]{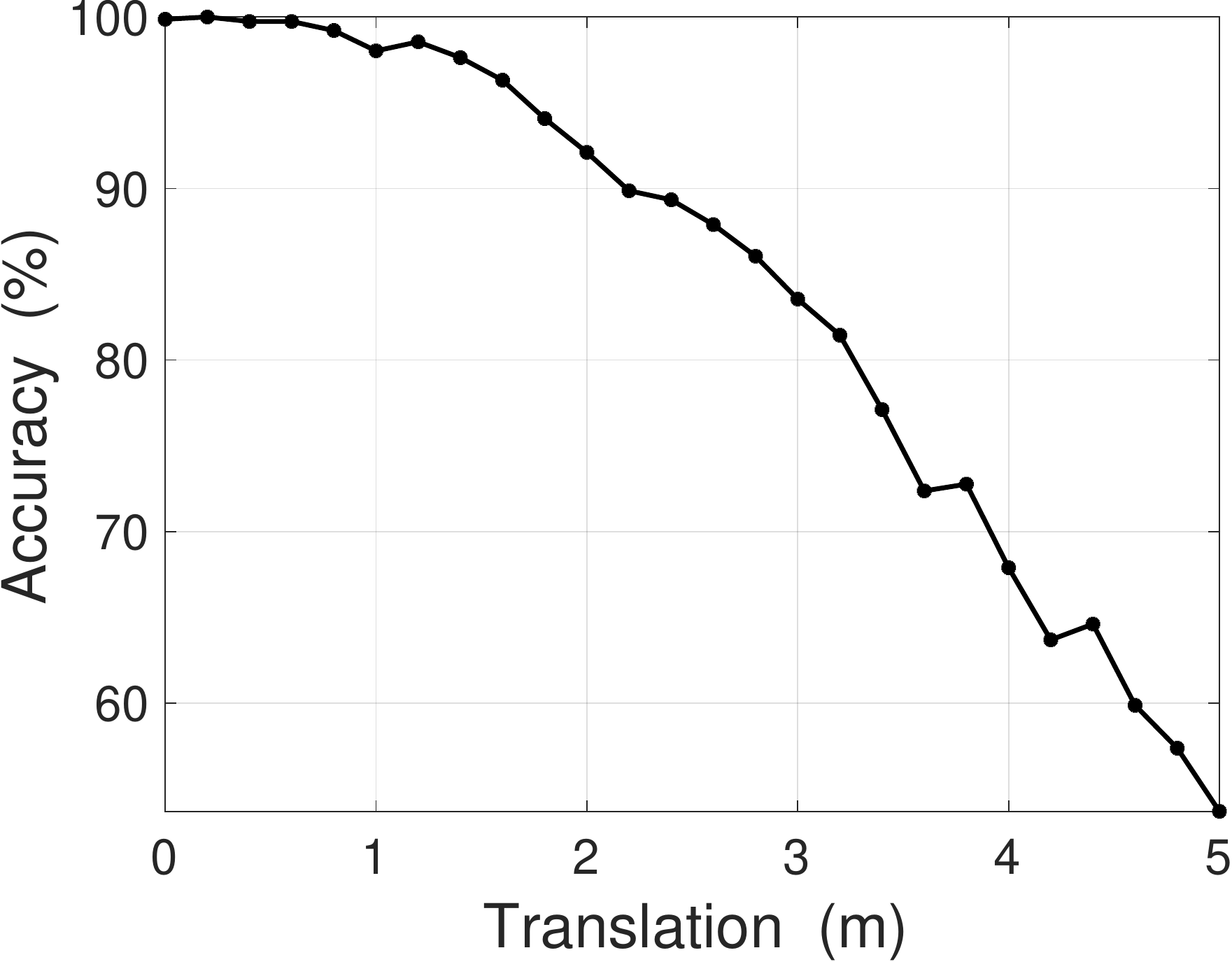} \vspace{-3pt}\\
    (a) & (b)
      \vspace{-10pt}
    \end{tabular}
  \caption{\small Effect of two types of noise on accuracy. (a) Effect of zero-mean Gaussian noise of variance $\sigma^2$, added to each point, and (b) effect of translations (in meters) to the right (simulated).}
  \label{fig:sigma}
  \vspace{-10pt}
\end{figure}

Fig.~\ref{fig:sigma} (a) shows the effect of adding zero-mean Gaussian random noise of variance $\sigma^2$ to each coordinate of the returns (point cloud) on accuracy. We notice that although the technique seems to be robust to lower levels of noise, the performance degrades with increasing $\sigma$. This is because the ``signatures" are heavily dependent on the relative position of objects in the environment. This is somewhat reassuring, it points to the fact that TensorMap is basing its decision on the relative placement of features, leveraged at the classification stage. 
\vspace{-8pt}

Next we study the effect of a second, perhaps more challenging type of noise: translations. Fig.~\ref{fig:sigma} (b) shows the effect of successively shifting the test sequence to the right on the accuracy (\%). We notice that the technique is successful up-to a translation of about $1$m, beyond which, the performance quickly degrades. Note that a similar effect can be observed for translations to the left. The translations we consider here are \textit{artificially} generated, in practice the effect of translation may be worse. This is because, the Lidar ``sees'' additional objects in the direction of translation; posing a potential challenge for our approach. 
\vspace{-10pt}
\subsection{Compression Ratio}
\vspace{-6pt}
Finally, we analyze the compression ratio of the proposed technique in terms of number of elements to be stored. For the given choice of parameters we achieve the ratio of TensorMap : Tensor representation : Lidar Scan representation of about 1 : 400 : 8300. \sr{This significant improvement in terms of memory requirement enables use of TensorMap in real-world applications.}
\vspace{-10pt}
\subsection{Other Applications and Future Work}
\vspace{-6pt}
Applications of TensorMap also include secure and efficient location communication by transmission of the ``signatures'' (which in the current case are just $5 \times 5$ matrices), these ``signatures'' can be viewed as encoded location information. These can be directly understood by the sender and receiver(s), who have access to the \textit{a priori} known topological map.  Further, as alluded to in Section~\ref{sec:intro}, TensorMap can be used for coarse localization before scan-matching thus reducing the associated computational and storage overhead, potentially making scan-matching viable for real-time localization. \sr{Further, TensorMap can also be used to detect false loop-closures while scan-matching.}

Future work includes fusing data from other sensors to improve the robustness of TensorMap in order to develop techniques for localization, and comparison of such a technique with related works. Also, as alluded to in this discussion, using unequal segment lengths, instead of the fixed ones considered here, remains a potential direction. 
  \vspace{-18pt}
\section{Conclusions}\label{conclusion}
  \vspace{-6pt}
\sr{Lidar scan-matching provides the most accurate information about the position of the autonomous vehicle, yet it is computationally expensive, prohibiting its use in real-time localization. Popular techniques reduce the rate of data acquisition to cope with this overhead. In this work, we present a technique based on tensor decompositions for building efficient (in terms of space complexity) graph representations of maps. 
Our preliminary investigation of the proposed technique via experimental evaluations on real-world Lidar data for a localization task shows promising results, and opens exciting avenues for future explorations, in order to make autonomous vehicle navigation safer and more stable.} 
\bibliographystyle{IEEEbib}
\bibliography{referLidar}

\begin{thebibliography}{10}

\bibitem{Pandey11}
G.~Pandey, J.~R. McBride, and R.~M. Eustice,
\newblock ``Ford campus vision and lidar data set,''
\newblock {\em International Journal of Robotics Research}, vol. 30, no. 13,
  pp. 1543--1552, 2011.

\bibitem{Siagian2009}
C.~Siagian and L.~Itti,
\newblock ``Biologically inspired mobile robot vision localization,''
\newblock {\em IEEE Transactions on Robotics}, vol. 25, no. 4, pp. 861--873,
  2009.

\bibitem{Wang2006}
J.~Wang, H.~Zha, and R.~Cipolla,
\newblock ``Coarse-to-fine vision-based localization by indexing
  scale-invariant features,''
\newblock {\em IEEE Transactions on Systems, Man, and Cybernetics, Part B:
  Cybernetics}, vol. 36, no. 2, pp. 413--422, 2006.

\bibitem{Fraundorfer2007}
F.~Fraundorfer, C.~Engels, and D.~Nist{\'e}r,
\newblock ``Topological mapping, localization and navigation using image
  collections,''
\newblock in {\em 2007 IEEE/RSJ International Conference on intelligent Robots
  and Systems (IROS)}. IEEE, 2007, pp. 3872--3877.

\bibitem{Booij2007}
O.~Booij, B.~Terwijn, Z.~Zivkovic, and B.~Kr{\"o}se,
\newblock ``Navigation using an appearance based topological map,''
\newblock in {\em 2007 IEEE International Conference on Robotics and
  Automation}. IEEE, 2007, pp. 3927--3932.

\bibitem{Chang2010}
C.~K. Chang, C.~Siagian, and L.~Itti,
\newblock ``Mobile robot vision navigation \& localization using gist and
  saliency,''
\newblock in {\em 2010 IEEE/RSJ International Conference on Intelligent Robots
  and Systems (IROS)}. IEEE, 2010, pp. 4147--4154.

\bibitem{Milford2012}
M.~J. Milford and G.~F. Wyeth,
\newblock ``Seqslam: Visual route-based navigation for sunny summer days and
  stormy winter nights,''
\newblock in {\em 2012 IEEE International Conference on Robotics and Automation
  (ICRA)}. IEEE, 2012, pp. 1643--1649.

\bibitem{Schindler2007}
G.~Schindler, M.~Brown, and R.~Szeliski,
\newblock ``City-scale location recognition,''
\newblock in {\em 2007 IEEE Conference on Computer Vision and Pattern
  Recognition (CVPR)}. IEEE, 2007, pp. 1--7.

\bibitem{Angeli2009}
A.~Angeli, S.~Doncieux, J.~A. Meyer, and D.~Filliat,
\newblock ``Visual topological slam and global localization,''
\newblock in {\em 2009 IEEE International Conference on Robotics and Automation
  (ICRA)}. IEEE, 2009, pp. 4300--4305.

\bibitem{Besl92}
P.~J. Besl and N.~D. McKay,
\newblock ``Method for registration of 3-d shapes,''
\newblock in {\em Robotics-DL tentative}. International Society for Optics and
  Photonics, 1992, pp. 586--606.

\bibitem{Biber03}
P.~Biber and W.~Stra{\ss}er,
\newblock ``The normal distributions transform: A new approach to laser scan
  matching,''
\newblock in {\em 2003 IEEE International Conference on Intelligent Robots and
  Systems (IROS)}. IEEE, 2003, vol.~3, pp. 2743--2748.

\bibitem{Morris2005}
A.~Morris, D.~Silver, D.~Ferguson, and S.~Thayer,
\newblock ``Towards topological exploration of abandoned mines,''
\newblock in {\em Proceedings of the 2005 IEEE International Conference on
  Robotics and Automation (ICRA)}. IEEE, 2005, pp. 2117--2123.

\bibitem{Myronenko10}
A.~Myronenko and X.~Song,
\newblock ``Point set registration: {C}oherent point drift,''
\newblock {\em IEEE Transactions on Pattern Analysis and Machine Intelligence},
  vol. 32, no. 12, pp. 2262--2275, 2010.

\bibitem{Mueller2011}
A.~Mueller, M.~Himmelsbach, T.~Luettel, F.~V. Hundelshausen, and H.~J.
  Wuensche,
\newblock ``Gis-based topological robot localization through lidar crossroad
  detection,''
\newblock in {\em 2011 14th International IEEE Conference on Intelligent
  Transportation Systems (ITSC)}. IEEE, 2011, pp. 2001--2008.

\bibitem{zhang14}
J.~Zhang and S.~Singh,
\newblock ``Loam: Lidar odometry and mapping in real-time,''
\newblock in {\em Robotics: Science and Systems Conference (RSS)}, 2014, pp.
  109--111.

\bibitem{zhang15}
J.~Zhang and S.~Singh,
\newblock ``Visual-lidar odometry and mapping: Low-drift, robust, and fast,''
\newblock in {\em IEEE International Conference on Robotics and Automation
  (ICRA)}. IEEE, 2015, pp. 2174--2181.

\bibitem{Tucker66}
L.~R. Tucker and R.~Ledyard,
\newblock ``Some mathematical notes on three-mode factor analysis,''
\newblock {\em Psychometrika}, vol. 31, no. 3, pp. 279--311, 1966.

\bibitem{De2000}
L.~De Lathauwer, B.~De Moor, and J.~Vandewalle,
\newblock ``A multilinear singular value decomposition,''
\newblock {\em SIAM journal on Matrix Analysis and Applications}, vol. 21, no.
  4, pp. 1253--1278, 2000.

\bibitem{Kolda09}
T.~G. Kolda and B.W. Bader,
\newblock ``Tensor decompositions and applications,''
\newblock {\em SIAM review}, vol. 51, no. 3, pp. 455--500, 2009.

\bibitem{Sidiropoulos2017}
N.~D. Sidiropoulos, L.~De Lathauwer, X.~Fu, K.~Huang, E.~E. Papalexakis, and
  C.~Faloutsos,
\newblock ``Tensor decomposition for signal processing and machine learning,''
\newblock {\em IEEE Transactions on Signal Processing}, vol. 65, no. 13, pp.
  3551--3582, July 2017.

\bibitem{Li2017}
N.~Li, N.~Pfeifer, and C.~Liu,
\newblock ``Tensor-based sparse representation classification for urban
  airborne lidar points,''
\newblock {\em Remote Sensing}, vol. 9, no. 12, 2017.

\end{thebibliography}

\end{document}